\documentclass[12pt]{article}
\usepackage [utf8]{inputenc}
\usepackage [english]{babel}
\usepackage{graphicx} 
\usepackage[top=1.0in, left=1.0in, bottom=1.0in, right=1.0in]{geometry}
\usepackage{mathtools}
\usepackage{amsmath}  
\usepackage{cancel}
\usepackage{slashed}
\usepackage{misccorr}
\usepackage{siunitx}
\usepackage{amsthm}
\usepackage{amssymb}
\usepackage{wrapfig}
\usepackage{booktabs}
\usepackage{enumitem}
\usepackage[vcentermath]{youngtab}
\usepackage{lmodern}
\usepackage{subcaption}
\usepackage{xcolor, soul}

\newtheorem{pr}{Proposition}[section]
\newtheorem{ex}{Example}[section]
\newtheorem{exer}{Exercise}[section]
\newtheorem{defn}{Definition}[section]
\newtheorem{rem}{Remark}[section]
\newtheorem{hint}{Hint}[section]
\newtheorem{cor}{Corollary}[section]
\newtheorem{thm}{Theorem}[section]
\newtheorem{nt}{Notation}[section]
\newtheorem*{problem}{\problemname}
\newcommand{\problemname}{}
\newenvironment{statement}[1]
 {\renewcommand{\problemname}{#1}\problem}
 {\endproblem}

\usepackage{hyperref}
\hypersetup{
    colorlinks=true,
    linktoc=all,    
    linkcolor=blue, 
    citecolor=red,
    urlcolor=black,
    anchorcolor = blue,
}

\numberwithin{equation}{section}
\newcommand\sdtimes{\mathbin{\hbox{\hskip2pt\vrule height 4.1pt depth -.3pt
width.25pt\hskip-2pt$\times$}}}
\newcommand{\ve}{{\varepsilon}}
\newcommand{\bve}{\boldsymbol{\varepsilon}}
\newcommand{\ep}{{\epsilon}}
\renewcommand{\Im}{{\operatorname{Im}\,}}

\newcommand\beq{\begin{equation}}
\newcommand\eeq{\end{equation}}

\theoremstyle{definition}

\newenvironment{aside}{\begin{quote}\sffamily}{\end{quote}}

\newcommand{\Ker}{{\mathrm{Ker}}}

\newcommand {\BC}   {\mathbb C}

\newcommand {\BR}   {\mathbb R}
\newcommand {\BP}   {\mathbb P}

\newcommand {\BT}   {\mathbb T}

\newcommand {\bR}   {\mathbf{R}}

\newcommand {\qe} {\mathfrak q}

\newcommand {\BZ}   {\mathbb Z}

\newcommand {\CalA} {\mathcal A}

\newcommand {\CalD} {\mathcal D}
\newcommand {\CalE} {\mathcal E}

\newcommand {\CalG} {\mathcal G}
\newcommand {\CalH} {\mathcal H}
\newcommand {\CalI} {\mathcal I}

\newcommand {\CalL} {\mathcal L}

\newcommand {\CalN} {\mathcal N}
\newcommand {\CalO} {\mathcal O}
\newcommand {\CalP} {\mathcal P}

\newcommand {\CalT} {\mathcal T}

\newcommand {\CalV} {\mathcal V}

\newcommand {\CalW} {\mathcal W}

\newcommand {\zb} {{\bar z}}

\newcommand {\fg} {\mathfrak{g}}

\newcommand{\fo}{\vert\kern -.03in\_}

\newcommand {\ii} {\mathrm{i}}
\newcommand{\Tr}{\text{Tr}}

\newcommand{\myhref}[2]{\hyperref[#1]{#2}}%

\title{Gauge theories and  Many-Body systems}

\author{Nikita Nekrasov, Igor Chaban\\
{\small
$^{\sf NN}$ Simons Center for Geometry and Physics, Stony Brook University, Stony Brook NY 11794}\\
{\small
$^{\sf IC}$ Department of Mathematics, Columbia University, New York NY 10027
}}

\date{\vspace{-5ex}}

\begin{document}
\maketitle
\tableofcontents
\newpage
\section{Introduction}

These lectures discuss two correspondences between gauge theories and integrable many-body systems. 
The first correspondence, based on infinite-dimensional Hamiltonian reduction and its quantum counterparts, grew out of the study of Calogero--Moser-type systems, two-dimensional Yang--Mills theory, and symplectic reduction in the 1980--1990s; see for example \cite{Calogero1971,Sutherland1971,Migdal:1975,Meyer1973,MARSDEN1974121,KKS78}. In this approach the quantization parameters of the gauge theory coincide with the quantization parameters of the many-body system. 
The second correspondence started emerging in the mid-1990s and later acquired a concrete non-perturbative realization in supersymmetric gauge theory and instanton counting; see \cite{Nekrasov:2002qd,Nekrasov2015BPSCFTCN}. It involves non-trivial dualities, relating classical problems on one side to quantum ones on the other and vice versa. This duality has various reincarnations: Fourier and Legendre transforms, Langlands duality, and more. The quantization parameters are mapped to geometric parameters. Simple questions on one side solve complicated ones on the other and vice versa. 

Why many-body systems? Gauge theories describe dynamics of gauge equivalence classes of connections, and the simplest observables to define in the classical gauge-theoretic setting 
are the conjugacy classes of the holonomies of the dynamical gauge field along a fixed loop in space time. If this loop is moved from one location to another, the conjugacy class of the holonomy changes. For classical gauge groups these conjugacy classes are in one-to-one correspondence with configurations of points on a circle modulo permutations and, possibly, other symmetries; thus the change in conjugacy class described above becomes the dynamics of indistinguishable point particles on a circle. 

In the quantum context, the positions of these particles become random variables. Depending on the dimensionality of spacetime and the degree of supersymmetry of the underlying gauge theory, the mapping of spacetime gauge dynamics to that of a many-body system could be exact or approximate. 

In what follows we study the Calogero--Moser--Sutherland class of
many-body systems, associated with $A$ type root system
and $SU(N)$ gauge theories in various spacetime dimensions, starting in $0+1$ and going all the way to $5+1$, with up to eight supercharges. The correspondence will
be direct in $0+1, 1+1, 2+1, 3+1$ dimensions for non-supersymmetric
matrix quantum mechanics, two-dimensional Yang--Mills
theory, modified three-dimensional Chern--Simons theory, 
or a hybrid 3d version of 2d Yang--Mills theory, and the hybrid
4d Chern--Simons theory. The correspondence will
be of the second type for $3+1, 4+1, 5+1$ dimensional
super-Yang--Mills theories with eight supercharges, subject to
the so-called $\Omega$-deformation. 

We shall introduce a class of non-local observables, 
generalizing the loop operator of the old-fashioned matrix models. Specifically for supersymmetric gauge theories the correlation functions of these observables obey a non-perturbative version of the Dyson--Schwinger equations that we review. From these equations, in the $3+1$ dimensional case, follows the Schr{\"o}dinger equation of a many-body system. The correspondence in the $4+1$ and $5+1$ dimensional cases remains a conjecture. 

The first part of these notes develops the symplectic-reduction construction of rational, trigonometric, and elliptic Calogero--Moser systems from gauge-theoretic data. The second part turns to supersymmetric gauge theories, equivariant localization on instanton moduli spaces, measures on partitions, and order/disorder observables. This split mirrors the two correspondences described above.

{\bf Acknowledgements.} These notes are mostly based on the lectures given by one of the authors at 2024 Beijing Summer Workshop in Mathematics
and Mathematical Physics ``Integrable systems and algebraic geometry'', dedicated to the memory of Igor
Krichever, at BIMSA, and partly on the lectures at the 2024 Les Houches Summer School ``Quantum Geometry - Mathematical Methods for Gravity, Gauge Theories and Non-Perturbative Physics''. We thank the organizers of the Schools for their hospitality.

NN is indebted to Sasha Gorsky, Andrei Losev, Edward Frenkel, Andrei Okounkov and especially to Igor Krichever for numerous discussions over the years. 

IC is grateful to Andrei Okounkov and Aleksandr Artemev for useful discussions.

Both of us thank Anton Dzhamay for his stimulating interest in the project.

Research of NN was supported by NSF PHY Award 2310279 and by the Simons Collaboration ``Probabilistic Paths to QFT''. 
NN thanks Institut Mittag-Leffler (Stockholm), Uppsala University, IHES, and especially ICTS-TIFR (Bengaluru) for their hospitality during the preparation of the manuscript.

\section{Calogero--Moser many-body system}
The first part of the notes explains how Calogero--Moser-type systems arise from Hamiltonian group actions, moment maps, and symplectic reduction. We begin with the rational model, which already contains the main ideas in their simplest form: a Lax representation, a larger linear Hamiltonian system, and a reduction by a unitary group action. The trigonometric and elliptic systems will then appear as natural variations of this construction and as gauge-theoretic analogues of the same mechanism.
\subsection{Definition of the system}
Like any classical Hamiltonian system, the Calogero--Moser (CM) system (\cite{Calogero1971}) is a symplectic manifold $(\mathcal{P}, {\omega})$  called {\sf the phase space} and a function $H$ on $\mathcal{P}$, {\sf the Hamiltonian}.

\begin{defn}
\normalfont
\begin{itemize}[label=\raisebox{0.25ex}{\tiny$\bullet$}]
(Rational) Calogero--Moser system is the triple
\item
    Phase space $\mathcal{P} = \big( T^{*}(\mathbb{R}^{N} \setminus \Delta ) \big) / S_{N}$
   with standard coordinates $\{(p_i, x_i)\}_{i = 1}^{N}$ on $T^{*}\mathbb{R}^{N}$, where $\Delta = \{x_i = x_j \text{ for some } i \neq j\}$ is the  union of diagonals and $S_{N}$ is the symmetric group acting on $(x_i, p_i)$ by permutations.
 \item
   The symplectic structure is defined by a canonical symplectic form
   \begin{align*}
       \omega = \sum \limits_{i = 1}^{N} d p_i \wedge dx_{i}.
   \end{align*}
   \item
  
    The CM Hamiltonian is given by the formula
    \begin{align*}
        H = \sum\limits_{i = 1}^{N} \frac{p_i^2}{2} + \nu^2 \sum\limits_{i < j} \frac{1}{(x_i - x_j)^2},
    \end{align*}
    where $\nu$ is the \textit{coupling constant} parameter.
\end{itemize}
\end{defn}
    The system describes motion of particles (many bodies) strung on a thread and interacting pairwise with potential $V(r) = \frac{\nu^2}{r^2}$ (repulsion for $\nu \in \mathbb{R}$).
\subsection{Integrals of motions}
\begin{exer}\normalfont
 Let us recall that the time evolution of a Hamiltonian system is described by a map $\gamma: \mathbb{R} \rightarrow \mathcal{P}$, satisfying equations of motion
 \begin{align*}
     \dot \gamma = V_{H},
 \end{align*}
where $V_{H}$ is the Hamiltonian vector field
\begin{align*}
    \omega(V_{H}, \cdot) = -dH,
\end{align*}
corresponding to the Hamiltonian $H$.\\
Show that the equations of motion for the Calogero--Moser system are presented in the form:
 \beq
     \begin{dcases}
         \dot x_{i} = \frac{\partial H}{\partial p_i} = p_i\\
         \dot p_i = - \frac{\partial H}{\partial x_i} = 2 \nu^2 \sum\limits_{k \neq i} \frac{1}{(x_i - x_k)^3}.
     \end{dcases}
     \label{eq:EOMcm}
 \eeq
 \end{exer}
 One can easily observe that total momentum $P = \sum \limits_{i = 1}^{N} p_i$  is a constant of motion. There are in fact $N$ algebraically independent integrals of motion that are Poisson commuting. When such exist the system is called integrable. Anyway the rest of $N - 2$ besides $H, P$ are harder to identify. A convenient way of exhibiting integrability is to find the \textit{Lax pair} $(L, A)$.
\begin{statement}{Problem 1}\normalfont\label{prob1}
    Consider two matrix-valued functions $L$ and $A$ on $\CalP$:
    \beq 
    L \, =\,  \Vert L_{ij} ({\bf p},{\bf x}) \Vert_{i, j = 1}^{N}\, , \ A \, = \, \Vert A_{i, j} ({\bf p},{\bf x}) \Vert_{i, j = 1}^{N} \in \mathrm{Mat}_{\mathbb{C}}(N)\, , 
    \eeq
    where 
 \begin{align}
     L_{ij} ({\bf p},{\bf x})  &= p_i \delta_{ij} + (1 - \delta_{ij})\frac{{\ii} \nu}{x_i - x_j}, \\
     A_{ij} ({\bf p},{\bf x})  &= (1 - \delta_{ij})\frac{{\ii} \nu }{(x_i - x_j)^2} - \delta_{ij}  \sum\limits_{k \neq i} \frac{{\ii} \nu}{(x_i - x_k)^2}. \label{Amat}
 \end{align}
Prove that the equations of motion \eqref{eq:EOMcm} are equivalent to the Lax equation 
\beq
\dot L = [A, L]\ .
\label{eq:laxeq}
\eeq
 \end{statement}
 From the Lax representation we immediately conclude that eigenvalues of $L$ are the integrals of motions. They have a clear mechanical interpretation of asymptotic momenta 
 \beq
 \left( p_1^{\pm}, ..., p_{N}^{\pm} \right) = {\rm Lim}_{t \rightarrow \pm \infty} \left( p_{1}(t) \,, \ldots \, , p_{N}(t) \right)\ .
\eeq
 Perhaps a more convenient way to parametrize the integrals of motion is the following:
 \begin{align*}
     H_{k} = \frac{1}{k} \mathrm{Tr} L^{k}
 \end{align*}
 \begin{rem}\normalfont
     Note that $H_1 = P, H_2 = H$.
 \end{rem}
 The functions $H_{k}$ for $k=1, ..., N$ are algebraically independent. Indeed, in the asymptotic region they become the power sums of the asymptotic momenta $p_1^{\pm}, \ldots, p_N^{\pm}$, and these power sums are algebraically independent on a Zariski-open set (equivalently, their Jacobian is a Vandermonde determinant). A direct check of Poisson commutativity of $H_k$'s is quite difficult. In the following we introduce a larger (in the sense of degrees of freedom) but simpler (in the sense of finding the solutions of EOMs) system. This will lead us to a proof of Poisson commutativity of $H_{k}$'s.
 \subsection{A simple system}
 Let us consider the following large but apparently simple system:
 \begin{defn}\normalfont\label{simpls}
 Let $k = 1, ..., N$. The Hamiltonian system $(\mathcal{P}_{L}, \omega_{L}, \boldsymbol{\varkappa})$ is the triple:
 \begin{itemize}[label=\raisebox{0.25ex}{\tiny$\bullet$}]
     \item 
     Phase space $\mathcal{P}_{L} = T^{*}(u(N) \times \mathbb{R}^{N})$. Here $u(N)=\mathrm{Lie}(U(N))$ is the Lie algebra of the unitary group. We choose the coordinates
     \begin{align*}
         &T^{*}u(N)= \big\{ \big( {\ii} P, {\ii} X \big) : P, X \in \mathrm{Mat}_{\mathbb{C}}(N), P^{\dagger} = P, X^{\dagger} = X \big\}\\
         &T^{*}\mathbb{R}^{N} = \big\{z = v + {\ii} w: v, w \in \mathbb{R}^N \big\}
     \end{align*}
     \item 
     Symplectic form $\omega_{L}$ is the standard one:
     \begin{align*}
         \omega_{L} = \mathrm{Tr}\big( dP \wedge dX \big) - {\ii}dz^{\dagger} \wedge dz.
     \end{align*}
     Here $dP \wedge dX$ is a matrix with entries being the $2$-forms defined by the matrix multiplication rule $\left( dP \wedge dX\right)_{ij} = \sum\limits_{k = 1}^{N}dP_{ik} \wedge dX_{kj}$. Similarly, $dz^{\dagger} \wedge dz = \sum\limits_{k = 1}^{N}dz^{\dagger}_{k} \wedge dz_{k}$, where $z_{k}^{\dagger} = {\bar z}^{k}$. 
     \item
     The Hamiltonians $\boldsymbol{\varkappa} = \left( {\varkappa}_{k} \right)_{k=1}^{N}$ are given by the formula
     \begin{align*}
         \varkappa_k = \frac{1}{k} \mathrm{Tr} P^k
     \end{align*}
 \end{itemize}
  \end{defn}
 \begin{exer}\normalfont
     Let $t_{k}$ be a time variable corresponding to the Hamiltonian $\varkappa_{k}$. Check that the equations of motion are given by
\begin{align*}
    \begin{dcases}
        \partial_{t_{k}}X_{ij} = \frac{\partial \varkappa_k}{\partial P_{ij}} = P_{ij}^{k-1}\\
        \partial_{t_k}P_{ij} = -\frac{\partial \varkappa_k}{\partial X_{ij}} = 0\\
        \partial_{t_k} z_{i} = {\ii}\frac{\partial \varkappa}{\partial \overline{z}_{i}} = 0,
    \end{dcases}
\end{align*}
and show that their solutions have a general form:
     \begin{align}\label{LEOM}
     \begin{cases}
         X(t) = X(0) + t_k P^{k - 1}(0)\\
         P(t) = P(0)\\
         z(t) = z(0).
     \end{cases}
     \end{align}
 \end{exer}
 \subsection{Hamiltonian action}
 A useful symmetry in the symplectic world is the \textit{Hamiltonian group action}.
 \begin{defn}\normalfont
Let $(M, \omega)$ be a symplectic manifold equipped with a smooth action of a Lie group $G$. The action is said to be Hamiltonian if there is a \textit{moment map}
\begin{align*}
    \mu : M \rightarrow \mathrm{Lie}^{*}(G),
\end{align*}
satisfying the conditions:
\begin{enumerate}
     \item[1.]
     for every $\xi \in \mathrm{Lie} (G)$, let $V_{\xi}$ denote the vector field generated by the one-parameter subgroup $\exp(t\xi)$. Then $V_{\xi}$ is the Hamiltonian vector field $V_{\xi} = V_{H_{\xi}}$:
   \beq
       d H_{\xi} = -\omega(V_{\xi}, \cdot)
\label{eq:hamxi}   \eeq     
with a Hamiltonian 
\begin{align}\label{Hxi}
    H_{\xi}(x) = \langle \mu(x), \xi \rangle.
\end{align}
\item[2.]
the $G-$\textit{equivariance} condition: for $\xi_1, \xi_2 \in \mathrm{Lie}(G)$ 
\beq
\{H_{\xi_1}, H_{\xi_2}\} = H_{[\xi_1, \xi_2]} 
 .
 \label{eq:mommappb}
 \eeq
\end{enumerate}
 \end{defn}
 Every Hamiltonian action on a symplectic manifold preserves the symplectic form. The converse is not true. If $M$ has nontrivial $H^{1}(M, {\BR})$, then there are vector fields $V$, which preserve the symplectic form, but do not have a Hamiltonian. Also, if the Lie algebra ${\mathrm{Lie}}(G)$ has nontrivial $H^{2}({\mathrm{Lie}}(G), {\BR})$, then even if there exist the Hamiltonians $H_{\xi}$ for each $\xi \in \mathrm{Lie}(G)$ as in \eqref{eq:hamxi}
 their Poisson brackets don't obey \eqref{eq:mommappb}, instead
 \beq
 \{ H_{\xi_1}, H_{\xi_2} \} = H_{[{\xi}_1, {\xi}_{2}]} + C({\xi}_{1}, {\xi}_{2})
 \eeq
 with some constant $C({\xi}_{1}, {\xi}_{2}) = - ({\xi}_{2}, {\xi}_{1})$, representing a non-trivial $2$-cocycle on $\mathrm{Lie}(G)$.

 We also need a notion of Hamiltonian action in case $(M, \omega)$ is equipped with a Hamiltonian function.
\begin{defn}\normalfont
An action of a Lie group $G$ on a Hamiltonian system $(M, \omega, H)$ is said to be Hamiltonian if it is Hamiltonian for the symplectic manifold $(M, \omega)$ and the function $H$ is $G$-invariant.
\end{defn}
\begin{rem}\label{rem:poiscom}
    In such a situation $H$ Poisson-commutes with $H_{\xi}$ for any $\xi \in \mathrm{Lie}(G)$. Indeed, $V_{H_{\xi}}.H = V_{\xi}.H=0$ since $H$ is $G$-invariant.
\end{rem}
\begin{pr}
    The natural $U(N)-$action on $\mathcal{P}_{L}$
\begin{align}\label{Uaction}
     g ( P, X, z ) = ( g^{-1} P g, g^{-1} X g, g^{-1} z )
 \end{align}
\end{pr}
preserves $\omega_{L}$ and is Hamiltonian.
 \begin{exer}\normalfont
     Check that 
\begin{align}
 &\mu_{L} : \mathcal{P}_{L} \rightarrow u(N)^{*} \stackrel{Tr}{\simeq} u(N) = \{ M \in \mathrm{Mat}_{\mathbb{C}}(N): M^{\dagger} = - M \}\nonumber \\
     &\mu_{L} (P, X, z) = [P, X] - {\ii} z \otimes z^{\dagger}\label{muL}
\end{align}
is the moment map for the action (\ref{Uaction}).
\end{exer}
\begin{hint}\normalfont
Since the natural pairing $\langle \cdot, \cdot \rangle$ on $u(n)$ is given by $\mathrm{Tr}$ it follows that
\begin{align*}
    H_{\xi}(P, X, z) &= \mathrm{Tr} \big( \xi ([P, X] - {\ii} z \otimes z^{\dagger})\big).
\end{align*}
Next, it is useful to find an explicit formula for $V_{\xi}$. For this purpose introduce a group element $g_{t} = \mathrm{exp}(\xi t) = 1_{N} + t \xi + O(t^2)$ and use the formula
\begin{align*}
    V_{\xi} = \frac{d}{dt}\bigg|_{t=0}g_{t}(P, X, z).
\end{align*}
\end{hint}
\begin{rem}\normalfont
    Note that $\mu$ is not unique: the map
\begin{align}\label{mmap}
    \mu (P, X, z) = [P, X] + {\ii} ( z \otimes z^{\dagger} - \nu 1_{N})
\end{align}
also satisfies the moment map conditions.
\end{rem}
\subsection{Symplectic reduction}
For the discussion below it is useful to formulate the following
\begin{thm}\label{th:MMW}
    (Meyer \cite{Meyer1973}, Marsden-Weinstein \cite{MARSDEN1974121}, see also \cite{woodward2011}) Let $(M, \omega)$ be a symplectic manifold equipped with a Hamiltonian $G$-action with a moment map $\mu$. Suppose also that $G$ acts properly and freely on $\mu^{-1}(0)$. Then $\mu^{-1}(0)/G$ has a structure of a smooth manifold of dimension $\dim M - 2\dim G$ with a unique symplectic form $\omega'$ satisfying $i^{*}\omega = \pi^{*}\omega'$, where
    $i: \mu^{-1}(0) \rightarrow M$ and $\pi: \mu^{-1}(0) \rightarrow \mu^{-1}(0)/G$ are the inclusion and the projection respectively.
\end{thm}
This allows us to produce a new Hamiltonian system starting from a Hamiltonian system with a Hamiltonian action. The resulting construction is called symplectic reduction.
\begin{defn}\normalfont
    Let $(M, \omega, H)$ be a Hamiltonian system equipped with a Hamiltonian action of a Lie group $G$ with a moment map $\mu$. The Hamiltonian reduction $(M', \omega', H') = (M, \omega, H) /\mkern-5mu/ G$ is a Hamiltonian system consisting of a triple
    \begin{itemize}[label=\raisebox{0.25ex}{\tiny$\bullet$}]
    \item 
    Phase space $M' = M /\mkern-5mu/ G \stackrel{def}{=} \mu^{-1}(0)^{ss} / G$, where $\mu^{-1}(0)^{ss}$ (ss for semistable) is some open subset of $\mu^{-1}(0)$ on which $G$ acts properly and freely. In general this subset depends on additional data, which is called stability conditions. In case of (\ref{mmap})  
    \begin{align*}
        \mu^{-1}(0)^{ss} = \big\{ (P, X, z) \in \mu^{-1}(0): X \text{ has different eigenvalues}\big\}.
    \end{align*}
    \item 
    Symplectic structure $\omega'$ is uniquely determined by the identity $i^{*}\omega = \pi^{*}\omega'$ from Theorem~\ref{th:MMW}.
    \item 
    Hamiltonian $H'$ comes from restriction of $G$-invariant $H$ on $\mu^{-1}(0)^{ss}$.
\end{itemize}
\end{defn}
\begin{rem}
    In the example (\ref{mmap}) the action is proper since $G$ is compact. In addition, $0$ is a critical value of $\mu$ and thus $G$ acts locally free (i.e. all the stabilizers are finite). However, the action is not free which makes us to consider $\mu^{-1}(0)^{ss}$.
\end{rem}
\begin{pr}\label{tangenth}
    Hamiltonian vector field $V_{H}$ corresponding to $H$ is tangent to $\mu^{-1}(0)^{ss}$ and lifts $V_{H'}$, i. e. for any $m \in \mu^{-1}(0)^{ss}$ $\pi_{*}V_{H}\big|_{m} = V_{H'}\big|_{\pi(m)}$ (in particular $\pi_{*}V_{H}\big|_{m}$ doesn't depend on the choice of $m$ in a fiber of $\pi$).
\end{pr}
\begin{proof}
    Let $\{\xi_{i}\}_{i}$ be some basis of $\mathrm{Lie}(G)$. By Remark \ref{rem:poiscom} $V_{H}.H_{\xi} = \{H, H_{\xi}\} = 0$, hence $V_{H}$ is tangent to $\mu^{-1}(0)^{ss} \subset \mu^{-1}(0) = \cap_{i}H_{\xi_{i}}^{-1}(0)$. To prove the second statement we need to check that for any $u \in T_{\pi(m)}M'$ $\omega'(\pi_{*}V_{H}\big|_{m}, u) = -dH(u)$. Let $\tilde{u} \in T_{m}\mu^{-1}(0)^{ss}$ be any lifting of $u$. Using the notation for $i$ and $\pi$ as in the formulation of the Theorem \ref{th:MMW} we have
\begin{align*}
    &\omega'(\pi_{*}V_{H}\big|_{m}, u) = \omega'(\pi_{*}V_{H}\big|_{m}, \pi_{*}\tilde{u}) = (\pi^{*}\omega')(V_{H}\big|_{m},\tilde{u}) = (i^{*}\omega)(V_{H}\big|_{m},\tilde{u}) = -d(i^{*}H)(\tilde{u}) =\\
    &=-d(\pi^{*}H)(\tilde{u}) = -dH(\pi_{*}u)=-dH(u).
\end{align*}
\end{proof}
\subsection{Hamiltonians of the reduction}
Let $(\mathcal{P}_{S}, \omega_{S}, H_{k})$ be the symplectic reduction of the Hamiltonian system $(\mathcal{P}_{L}, \omega_{L}, \varkappa_{k})$ (\ref{simpls}). We define some natural functions on $\mathcal{P}_{L}$. Namely, $x_1, ..., x_N$ are the eigenvalues of $X$ and $p_1, ..., p_N$ are the diagonal entries of $P$.
Our aim is to find $H_k$ explicitly in terms of $(p_i, x_i)$ (see \cite{KKS78}).
\begin{rem}\normalfont
    The functions $x_1, ..., x_N$ are $U(N)-$invariant and hence give rise to the well defined functions on $\mathcal{P}_{S}$. $p_1 , ..., p_N$ are not $U(N)-$invariant.
\end{rem}
\begin{pr}
    $\mathcal{P}_{S}$ is symplectomorphic to $T^{*}\{(x_1 , ..., x_n ): x_1 < ... < x_n \}$.
\end{pr}
\begin{proof}
Let us smoothly pick a representative in every $U(N)-$orbit of $\mu^{-1}(0)^{ss}$. The set of representatives is a leaf transversal to $U(N)-$orbits and diffeomorphic to the quotient space $\mathcal{P}_{S}$. The functions $p_1, ..., p_N$ are thus defined on $\mathcal{P}_{S}$ in terms of restriction on the leaf. The choice of representatives (gauging) is our freedom. To simplify the calculations one should use this freedom wisely. We propose the following choice. In every $U(N)-$orbit we pick a diagonal $X = \mathrm{diag}(x_1, ..., x_N ), ~ x_1 < ... < x_N$.
The remaining symmetry is $\mathrm{Stab}_{U(N)}(X) = U(1) \times ... \times U(1) \subset U(N)$.
In the $U(1)^{\times N}-$orbit we choose $z_i \in \mathbb{R}_{+}$. This fixes the gauge completely. In fact due to the constraint $\mu = 0$ the coordinates $P_{ij}$ and $z_i$ are determined completely with a given $x_i$ and $p_i$ on the leaf chosen.
\begin{statement}{Problem 2}\normalfont\label{prob2}
Show that the conditions $\mu = 0$, $X = \mathrm{diag}(x_1, ..., x_n)$, $x_1 < ... <x_N$ , $z_i \in \mathbb{R}_{+}$ imply
\begin{align*}
\begin{dcases}
        z_i = \sqrt{\nu}, \;i = 1, ..., N\\\\
        P_{ij} = \frac{{\ii}\nu}{x_i - x_j}, \;i \neq j,
    \end{dcases} 
\end{align*}
the matrix $P$ hence has the form
\begin{align}\label{Pmatrix}
    P_{ij} = p_{i} \delta_{ij} + \frac{{\ii}\nu (1 - \delta_{ij})}{x_i - x_j}.
\end{align}
\end{statement}
Thus, identifying $p_i$ with a coordinates on the cotangent fibers we see that the leaf is diffeomorphic to $T^{*}\{(x_1 , ..., x_n ): x_1 < ... < x_n \}$. The following exercise identifies the symplectic forms and thereby concludes the proposition.
\begin{exer}\normalfont
    Using the defining property $\pi^{*}\omega_{S} = i^{*}\omega_{L}$ for $\omega_{S}$, where $i: \mu^{-1}(0)^{ss} \hookrightarrow \mathcal{P}_{L}$ and $\pi: \mu^{-1}(0)^{ss} \rightarrow \mu^{-1}(0)^{ss}/U(N)$, check that
    \begin{align*}
        \omega_{S} = \sum\limits_{i = 1}^{N} d p_{i} \wedge d x_{i}
    \end{align*}
\end{exer}
\end{proof}
\begin{cor}
   The reduction Hamiltonians are $H_{k} = \frac{1}{k}\mathrm{Tr} P^{k}$, where matrix $P$ is given by $(\ref{Pmatrix})$.
\end{cor}
\begin{rem}\normalfont
    Note that $\varkappa_{k}$'s obviously Poisson-commute. By (\ref{tangenth}) 
\begin{align*}
    \pi^{*}\{H_{k}, H_{l}\} = \pi^{*}(\omega_{S}(V_{H_{k}}, V_{H_{l}})) = (\pi^{*}\omega_{S})(v_{\varkappa_{k}}, v_{\varkappa_{l}})=
    \omega_{L}(v_{\varkappa_{k}}, v_{\varkappa_{l}}) = \{\varkappa_{k},\varkappa_{l}\}=0.
\end{align*}    
Since $\pi^{*}$ is injective this implies that $H_{k}$'s Poisson-commute as well.
\end{rem}
We see that the system produced by symplectic reduction is none other than the previously considered Calogero--Moser system with the Lax matrix $L = P$. The matrix $A$ also appears naturally. Let $(P, X, z) = (P(t), X(t), z(t))$ be the solution $(\ref{LEOM})$. Let $g = g(t)$ be the group element such that $(\tilde{P}, \tilde{X}, \tilde{z}) \overset{def}{=} (g^{-1} P g, g^{-1} X g, g^{-1}z)$ stays on the chosen leaf. Since $P(t) = P(0)$, we see $\dot{\tilde{P}} = [\tilde{P}, g^{-1} \dot g]$ and conclude
\begin{align*}
      A &= -g^{-1} \dot g.
\end{align*}
\begin{exer}
    Compute $g^{-1} \dot g$ for $t = t_{2}$ and compare it with the $A$ from the $(\ref{Amat})$.
\end{exer}
\section{Gauge theory counterpart}
In the following we give a gauge theory perspective on the subject. This provides a natural generalization of the Hamiltonian system discussed.  
\subsection{Lagrangian description}
Let us consider the action functional for the system $(\mathcal{P}_{L}, \omega_{L}, \varkappa)$ in the first-order formalism. We see that $\omega_{L} = d \alpha$ for $\alpha = \mathrm{Tr} P d X + {\ii} z^{\dagger} dz$. Thus
\begin{align*}
    S_{L} = \int \alpha - H dt = \int \mathrm{Tr} (P d X) + {\ii} z^{\dagger} d z - \frac{1}{2} \mathrm{Tr} P^2 dt
\end{align*}
Note that this action is not gauge invariant. One can impose the condition $\mu = 0$ by introducing a Lagrange multiplier ${\ii}A_{t} \in u(N)$.
\begin{align}
    S = \int \mathrm{Tr} (P d X) + {\ii} z^{\dagger} d z - \frac{1}{2} \mathrm{Tr} P^2 dt + \mathrm{Tr} (\mu (P, X, z) A_{t}) dt
\end{align}
\subsection{$2d$ Yang--Mills theory}
\begin{exer}\normalfont
    Show that for $S$ to be gauge invariant $A_{t}$ must transform
\begin{align*}
    {\ii} A_{t} \mapsto g^{-1} {\ii} A_t g - g^{-1} \dot g,
\end{align*}
which is transformation law of a connection one-form.
\end{exer}
We see that $S$ is the functional on
\begin{align*}
    \big( \mathrm{Maps}(\mathbb{R}, \mathcal{P}_{L}) \times G-\text{connections}\big)\big/ \mathrm{Maps}(\mathbb{R}, G).
\end{align*}
Let us now rewrite $S$ in the following form:
\begin{align*}
    S = \int \mathrm{Tr} \big( P [\partial_{t} - {\ii}A_t , X ] - \frac{1}{2}P^2 + (\partial_{t} - {\ii}A_t ) J \big)dt,
\end{align*}
where
\begin{align*}
    J = {\ii}(z \otimes z^{\dagger} - \nu 1_{N}).
\end{align*}
Gauge theory motivates replacing $X$ to ${\ii}\partial_{x} + A_x$. Such a generalization turns the connection $d - {\ii}A_{t} d t$ on $\mathbb{R} \ni t$ into the connection $D = d - {\ii}A_t d t - {\ii}A_x d x$ on $\mathbb{R} \times \mathbb{R} \ni (t, x)$ (or $\mathbb{R} \times S^{1}$). Analogous object in electrodynamics is called $\textit{vector potential}$. In this way $[\partial_{t} - {\ii}A_t , X]$ becomes a curvature form ${\ii}F_{tx}^{A} = {\ii}[\partial_{t} - {\ii}A_t , \partial_x - {\ii}A_x ]$.
Being a conjugated variable to the $X$ the quantity $P$ must be replaced with a function $E(x)$. Its electrodynamical analogue is $\textit{electric field}$. We want the term $J$ to impose some condition on $A_x$ at single point ($x = 0$) only (we will see later that this is reasonable), so let us replace it by $J \delta(x) dx$. The new action thus has the form
\begin{align*}
    S_{YM} = \int \mathrm{Tr} ({\ii} E F^{A}) - \frac{1}{2} \mathrm{Tr} (E^2) dx dt + \mathrm{Tr} (D_{t}^{A} J) \delta(x) dx dt.
\end{align*}
This is the $2d$ Yang--Mills theory (\cite{Migdal:1975}) in the first-order formalism. Let us rewrite it in the following way
\begin{align*}
    S_{YM} = \int \mathrm{Tr} \big( - \frac{1}{2} E^2 + D_{t} ({\ii}D_{x}E - J \delta(x)) \big) dx dt.
\end{align*}
\subsection{Hamiltonian description}
Instead of using the Lagrange multiplier $D_{t}$ we can simply consider the dynamics given by
\begin{align}\label{Sinf}
    S_{\infty} = - \int \frac{1}{2} \mathrm{Tr} ( E^2 ) dx dt
\end{align}
and impose the constraint $\mu_{\infty} \overset{def}{=} {\ii}D_x E - J \delta(x) = 0$ ($\textit{Gauss law}$).
\begin{rem}\normalfont
Note that $\mu_{\infty}$ is the moment map for the gauge group $\mathcal{G} = \mathrm{Maps}(S^1 \rightarrow U(N))$, which acts on $(E, A_{x}, z)$ by the rule
\begin{align*}
    g\big(E, A_x, z \big) = \big( g^{-1} E g,~ g^{-1} A_x g - g^{-1} \partial_x g , ~g^{-1}(0) z \big).
\end{align*}
\end{rem}
A new phase space formed by $(E(x), {\ii}\partial_{x} + A_{x} (x), z)$ is
\begin{align*}
    \mathcal{P}_{\infty} = T^{*} \big( u(N)-\text{connections on }S^1 \big) \times \mathbb{C}^{N}.
\end{align*}
The canonical symplectic form $\omega_{\infty}$ is given by
\begin{align*}
    \omega_{\infty}(f, \alpha) = \int\limits_{S^{1}} Tr(f \alpha ),
\end{align*}
where $\alpha \in T_{A_{x}}\big( u(N)-\text{connections on }S^{1} \big) \cong \Omega^{1}(S^{1}, u(N))$ and $f \in T_{E}\mathrm{Maps}(S^{1}, u(N)) \cong \mathrm{Maps}(S^{1}, u(N))$.
The Hamiltonian of the system $(\ref{Sinf})$
\begin{align*}
    \mathcal{H} = \frac{1}{2\pi} \int \limits_{0}^{2\pi} \frac{1}{2} \mathrm{Tr} (E^2) dx
\end{align*}
We again have a (very) large system $(\mathcal{P}_{\infty}, \omega_{\infty}, \mathcal{H})$ with Hamiltonian action of $\mathcal{G}$ that wants us to reduce it symplectically.
\subsection{Calogero--Moser--Sutherland system}
Analogously to the previous case we need to find a nice representative in every orbit of $\mathcal{G}$.
Every $u(N)$-connection $A_x$ on $S^1$ defines a parallel transport along the circle the monodromy 
matrix $M \in U(N)$. Gauge invariant eigenvalues $e^{2 \pi {\ii} x_1} , ..., e^{2 \pi {\ii} x_N } $ of $M$ provide natural set $(x_1, ..., x_N)$ of functions on $\mathcal{P}_{\infty}$ well defined on $\mathcal{P}_{\infty} /\mkern-5mu/ \mathcal{G}$ up to permutations and shifts:
\begin{align*}
    g = \mathrm{diag}(e^{{\ii} n_i x}) : (x_1, ..., x_N) \mapsto (x_1 + n_1 , ..., x_N + n_N ), ~n_i \in \mathbb{Z}.
\end{align*}
\begin{rem}\normalfont
    Note that such transformations form an affine Weyl group $S_N \ltimes \mathbb{Z}^{N}$ of $\widehat{u(N)}$. This is the analogue of the Weyl group $S_N$ of $u(N)$ acting by permutations on eigenvalues of $X$ in the finite-dimensional case.
\end{rem}

Using the Floquet--Lyapunov theory (\cite{Fl}, \cite{Lyap}) one can prove that every $A_x$ can be gauged to the constant matrix of the form $A_x = \mathrm{diag}(x_1, ..., x_N )$. The remaining symmetry is
\begin{align*}
    \mathrm{Stab}_{\mathcal{G}}(\mathrm{diag}(x_1, ..., x_N )) = \{ g(x) = \mathrm{diag} (e^{{\ii}\theta_{1}},...,e^{{\ii}\theta_{N}} ): \theta_i = \mathrm{const} \in \mathbb{R}) \} = U(1)^{\times N}.
\end{align*}
As before we use it to make $z_i \in \mathbb{R}_{+}$.
\begin{rem}\normalfont
Note that if $z_i$ were functions of $x$, the reduced phase space $\mathcal{P}_{S}$ would be infinite-dimensional (since the remaining gauge symmetry $U(1)^{\times N}$ has finite dimension). That is why in our generalization we left $z_i$ as constants rather than making them fields.
\end{rem}

Let $p_i$ be diagonal entries of $E(x = 0)$.

\begin{exer}\normalfont
    The solution of $\mu_{\infty} = 0$ in the chosen gauge is given by
\begin{align*}
E_{ij}(x) = 
\begin{dcases}
      p_{i}, \;\;i = j \\
     \frac{{\ii} \nu e^{-\pi {\ii} (x_i - x_j)}}{2 \sin (\pi (x_i - x_j))} e^{{\ii} (x_i - x_j) x}, \;\;i \neq j
\end{dcases}
\end{align*}
\end{exer}
It follows that the functions $(x_i, p_i)$ provide the coordinate system on the chosen leaf in $\mu_{\infty}^{-1}(0)^{ss}$.
\begin{exer}\normalfont
    Show that
    \begin{align*}
        \mathcal{H} = H_{2}^{tCM} \stackrel{def}{=} \sum\limits_{i = 1}^{N} \frac{p_i^2}{2} + \frac{\nu^2}{4} \sum\limits_{i < j} \frac{1}{\mathrm{sin}^{2}(\pi (x_i - x_j))}.
    \end{align*}
    This is the Hamiltonian of Calogero--Moser--Sutherland system (also called the trigonometric CM system) (\cite{Sutherland1971}, \cite{Moser1975}), derived from $2d$ Yang-Mills theory, following \cite{Gorsky:1993pe}.
\end{exer}
The construction above is an example of the promised direct correspondence between classical integrable systems and gauge theories. Here is a list of further examples:
\begin{itemize}[label=\raisebox{0.25ex}{\tiny$\bullet$}]
    \item 
    Elliptic Calogero--Moser corresponds to hybrid topological-holomorphic $2d$ Yang--Mills
    
    \item
    Relativistic rational Ruijsenaars--Schneider (\cite{Ruijsenaars:1986vq}, \cite{Ruijsenaars:1986pp}),
    \cite{Ruijsenaars:1988pv},
      \cite{RuijSim1990},
    \cite{Ruijsenaars:1994gc})
    corresponds to the topological $2d$ Yang-Mills, perturbed by a Wilson loop operator \cite{Fock:1999ae}. 
    
    \item 
    Trigonometric Ruijsenaars--Schneider corresponds to $3d$ Chern--Simons (\cite{Witten:1988hc})
    perturbed by a Wilson loop operator \cite{Gorsky:1993dq}
    
    \item 
    Elliptic Ruijsenaars--Schneider corresponds to hybrid topological-holomorphic $4d$ Chern--Simons (\cite{Nekrasov1996},\cite{costello2013})
\end{itemize}
In the following we consider the elliptic Calogero?Moser system.
\section{Elliptic Calogero--Moser system}
\begin{defn}\normalfont
    Elliptic Calogero--Moser system (\cite{Calogero:1975}) is a Hamiltonian system containing a triple
\begin{itemize}[label=\raisebox{0.25ex}{\tiny$\bullet$}]
\item Phase space of this system is a delicate matter discussed below. As a first approximation we take some open subset of $T^{*}\mathbb{C}^{N} = \{(p_i , x_i)_{i = 1}^{N}: p_i , x_i \in \mathbb{C}\}$. 
\item
    Hamiltonian is given by the formula
\begin{align*}
    H = \sum\limits_{i = 1}^{N} \frac{p_i^2}{2} + \nu^2 \sum\limits_{i < j} \wp (x_i - x_j , \tau),  
\end{align*}    
where
\begin{align*}
    \wp (x, \tau) =\mathrm{const} + \frac{1}{x^2} + \sum\limits_{(n, m) \in \mathbb{Z}^2 \setminus 0} \frac{1}{(x + m + n\tau)^2} - \frac{1}{(m + n \tau)^2}
\end{align*}
is the Weierstrass $\wp-$function on elliptic curve $C_{\tau} = \mathbb{C} \big/ \mathbb{Z} + \tau \mathbb{Z}$ and $\nu \in \mathbb{R}_{+}$ is the coupling constant.
\item The Poisson structure is given by the standard symplectic form
\begin{align*}
    \omega^{\mathbb{C}} = \sum\limits_{i = 1}^{N} dp_i \wedge dx_i
\end{align*}
\end{itemize}
\end{defn}
One can also consider real-valued $x_i$ and $p_i$. Analogously to the usual CM this system describes motion of particles interacting with a slightly more sophisticated potential. However allowing positions and momenta to be complex provides an interesting feature: points can swap without gaining infinite energy. This circumstance prompts us to consider an extended phase space. 
\subsection{Compactification of the phase space}
The actual phase space $\mathcal{P}^{\mathbb{C}}$ we are interested in is some compactification of $(T^{*}C_{\tau}^{\times N} \setminus \Delta ) \big/ S_N$, where $\Delta$ is the union of the diagonals  $\{x_i = x_j \}$.
\begin{rem}\normalfont
    The way $\omega^{\mathbb{C}}$ extends to compactification depends on $\nu$.
\end{rem}
Instead of giving a full definition, we illustrate the idea in the special case 
$N=2$. For simplicity, we restrict to the center-of-mass frame
   \begin{align*}
       \{x_1 + x_2 = 0, ~ p_1 + p_2 = 0\} \simeq T^{*}C_{\tau} \subset T^{*}(C_{\tau} \times C_{\tau})
   \end{align*}
In coordinates $x = \frac{x_1 - x_2}{2}, p = \frac{p_1 - p_2}{2}$ the group $\mathbb{Z}/2\mathbb{Z}$ acts on $T^{*}C_{\tau}$ by $(p, x) \mapsto (-p , -x)$.
The action has four fixed points $(p, x) = (0, 0),~(0, \frac{1}{2}),~(0, \frac{\tau}{2}),~(0, \frac{1 + \tau}{2})$.
\subsection{Around fixed points}
We need to define a natural compactification of $\big( T^{*}C_{\tau} \setminus \{\text{fixed points}\} \big) \big/ \mathbb{Z}/2\mathbb{Z}$.
The neighborhood of the fixed point $(0, 0)$ in $T^{*}C_{\tau} \big/ \mathbb{Z}/2\mathbb{Z}$ looks like
\begin{align*}
    \mathbb{C}^2 \big/ \mathbb{Z}/2\mathbb{Z} = \mathrm{Spec} (\mathbb{C}[z_1, z_2]^{\mathbb{Z}/2\mathbb{Z}}) = \mathrm{Spec} \big( \mathbb{C}[X, Y, Z] \big/ (X^2 - YZ) \big),
\end{align*}
where $X = z_1 z_2, Y = z_1^2, Z = z_2^2$.
The  singularity can be resolved (blowup) or deformed. Here we discuss its complex deformation, as  given by
\begin{align*}
    S_{\zeta} = \mathrm{Spec} \big( \mathbb{C}[X, Y, Z] \big/ (X^2 - YZ - \zeta^2) \big).
\end{align*}
For $\zeta \neq 0$ all $S_{\zeta}$ are isomorphic to each other. Moreover, in a rotated complex structure they can be identified with $\mathrm{Bl}_0 \big( \mathbb{C}^2 \big/ \mathbb{Z}/2\mathbb{Z} \big)$ -- the blow up $\mathbb{C}^2 \big/ \mathbb{Z}/2\mathbb{Z}$ at $(0, 0)$. $\zeta$ actually parameterizes the way $\omega^{\mathbb{C}}$ extends to compactification. More precisely, the parameter $\zeta$ is the period of the cohomology class $[\omega^{\mathbb{C}}]$ on the $2$-cycle, the generator of $H_{2}(S_{\zeta}, \mathbb{Z})$. This generator is the exceptional divisor $D$ in the blowup geometry $\mathrm{Bl}_0 \big( \mathbb{C}^2 \big/ \mathbb{Z}/2\mathbb{Z} \big)$, or the real two-sphere in the deformed geometry. To see this let us  use the parametrization
\begin{align*}
\begin{cases}
      X &= \zeta x\\
    Y &= \zeta (y + {\ii} z)\\
    Z &= -\zeta (y - {\ii} z)
\end{cases}
\end{align*}
with real $(x,y,z)$ obeying $x^2 + y^2 + z^2 = 1$.The symplectic 2-form $\omega^{\mathbb{C}}$ restricted on this cycle is given by
\begin{align*}
      \omega^{\mathbb{C}} = d z_1 \wedge d z_2 = \frac{d Y \wedge d Z}{4 X} = {\ii} \zeta \frac{dy \wedge dz}{2x}.
\end{align*}
The period is easy to compute now:
\begin{align*}
    \langle [\omega^{\mathbb{C}}], [D] \rangle =  \langle [\omega^{\mathbb{C}}], [S^2] \rangle =
    \int\limits_{S^2} \omega^{\mathbb{C}} = {\ii} \zeta \int\limits_{S^2}\frac{dy \wedge dz}{2x} = {\ii} \zeta \int\limits_{S^2}\mathrm{Vol}_{S^2} = 4 \pi {\ii} \zeta.
\end{align*}

The surface $\mathcal{P}^{\mathbb{C}}$ is $(T^{*}C_{\tau}) \big/ \mathbb{Z}/2\mathbb{Z}$ blown up at 4 points. A particular ($\nu-$dependent) choice of extension of the $\omega^{\mathbb{C}}$ to compactification is relevant for our system.
\subsection{Phase space as an algebraic variety}
Now let us find the explicit algebraic equation defining $(T^{*}C_{\tau})\big/ \mathbb{Z}/2\mathbb{Z}$. Consider three $\mathbb{Z}/2\mathbb{Z}-$invariant functions on $T^{*}C_{\tau}$:
\begin{align*}
    A = \wp(x), ~~ B = \frac{p \wp'(x)}{2 \nu}, ~~ u = \frac{H}{\nu} = \frac{p^2}{\nu} + \wp(x)
\end{align*}
Using the well-known identity
\begin{align}\label{efi}
    \wp'^2 = 4 \wp^3 - g_2(\tau) \wp - g_3 (\tau)
\end{align}
we find $B^2 = (u - A)(A - e_1)(A - e_2)(A - e_3)$, where $e_1, e_2, e_3$ are the roots of the polynomial in the right hand side of $(\ref{efi})$ (just some $\tau-$dependent constants).
We thus come up to the
\begin{pr}
    $T^{*}C_{\tau} \big/ \mathbb{Z}/2\mathbb{Z}$ is a fibration over the line $\mathbb{P}^1 \ni u$ with fibers $\mathcal{E}_{u}$ being elliptic curves. The four special fibers corresponding to $u = e_1, e_2, e_3, \infty$ are the degenerated curves with nodal singularities. A neighborhood of each of them in the compactified surface is modeled by $\mathrm{Spec}\big( [X, Y, Z]/(X^2 - YZ) \big)$.
\end{pr}
\section{Complexification of $tCM$ -- $2d$ Yang--Mills correspondence}
In the following we describe the complex analogue of the infinite-dimensional Hamiltonian system considered above. The symplectic reduction in this setting will reproduce the elliptic CM system. By complexification we mean passage from left to the right in the following table.
\newpage
\begin{table}[h]
   \label{tab:example}
   \small
   \centering
   \begin{tabular}{l|c|c}
    & \textbf{Real} & \textbf{Complex} \\ 
   \midrule
Gauge group & $G = U(N)$ & $G^{\mathbb{C}} = GL(N, \mathbb{C})$\\
\midrule
Lie algebra & $\mathfrak{g} = u(N)$ & $\mathfrak{g}^{\mathbb{C}} = gl(N, \mathbb{C})$ \\
\midrule
Source space & $M = (S^1 = \mathbb{R}/2 \pi \mathbb{Z} , x = 0)$  & $M = (C_{\tau} = \mathbb{C} / \mathbb{Z} + \tau \mathbb{Z}, z = 0)$  \\
&  -- pointed circle & -- pointed elliptic curve\\
\midrule
Phase space & $\mathcal{P}_{\infty} = \{ (E, A_x): A_x \in \mathrm{Conn}(S^1, \mathfrak{g}),  $ & $\mathcal{P}_{\infty}^{\mathbb{C}} = \{ (E, \bar{A}): \bar{A} \in \mathrm{Conn}^{0, 1}(C_{\tau}, \mathfrak{g}^{\mathbb{C}}), $\\
&$E \in \mathrm{Maps}(S^1 , \mathfrak{g})\} \times \mathbb{C}^N$ & $E \in \Omega^{1, 0}(C_{\tau}, \mathfrak{g}^{\mathbb{C}})\} \times \mathbb{C}^N \times (\mathbb{C}^N)^*$\\
\midrule
Hamiltonian & $H =\frac{1}{2\pi} \int \limits_{0}^{2\pi} \frac{1}{2} \mathrm{Tr} E^2 dx$ & $H =\frac{1}{\Im \tau} v.p. \int\limits_{C_{\tau}} \frac{1}{2} \mathrm{Tr} E^2 dz d \bar{z}$ \\
\midrule
Symmetry & $\mathcal{G} = \mathrm{Maps}(S^1, G)$ & $\mathcal{G}^{\mathbb{C}} = \mathrm{Maps}(C_{\tau}, G^{\mathbb{C}})$ \\
& $g(x)(E, A_x, z) = $ & $g(z, \bar{z}) (E, A, v, w) = $ \\
&$(g^{-1} E g, g^{-1} A_{x} g - {\ii} g^{-1} \partial_{x} g, g^{-1}(0)z)$&$(g^{-1} E g, g^{-1} \bar{A} g - g^{-1} \partial_{\bar{z}} g, g^{-1}(0) v, g(0)w )$\\
\midrule
Moment map &$\mu_{\infty} = [\partial_{x} - A_x , E] + J \delta (x)$& $\mu_{\infty}^{\mathbb{C}} = [\partial_{\bar{z}} - \bar{A}, E] + J^{\mathbb{C}} \cdot 2\pi \delta^{2}(z, \bar{z})$\\
&$J = {\ii} (z \otimes z^{\dagger} - \nu 1_{N})$& $J^{\mathbb{C}} = {\ii} (\nu 1_{N} - v \otimes w^{t})$
   \end{tabular}
\end{table}
Analogously to the real construction we want to find a nice representative in every orbit of $\mathcal{G}^{\mathbb{C}}$. We used the Floquet--Lyapunov theory before to establish the possibility of gauging the connection matrix $A_x$ to a certain form. In the complexified case we need the following statement:
\begin{thm}
(\cite{Atiyah1957VectorBO}) General rank $N$ complex vector bundle of degree $0$ on $C_{\tau}$ is isomorphic to $\bigoplus\limits_{i = 1}^{N} L_i $, where each $L_i$ is a degree $0$ line bundle.
\end{thm}
\subsection{Complex gauging}
It follows that general $\bar{A}$ is gauge equivalent to $\mathrm{diag}(\alpha_1 , ... , \alpha_N), \; \alpha_{i} \in \mathbb{C} $. It is convenient to rescale $\alpha_i$ introducing the variables $x_i = \frac{\Im \tau}{\pi} \alpha_i$. The functions $ x_1 , ..., x_N$ on $\mathcal{P}_{\infty}^{\mathbb{C}}$ are well-defined on $\mathcal{P}_{\infty}^{\mathbb{C}} 
/\mkern-5mu/ \mathcal{G}^{\mathbb{C}}$ up to permutation and shifts
\begin{align*}
    g = \mathrm{diag}(e^{\frac{2 \pi {\ii}}{\tau - \bar{\tau}}((z - \bar{z})n_i + (z \bar{\tau} - \bar{z} \tau)m_i)}): x_i \mapsto x_i + m_i + n_i \tau.
\end{align*}
The remaining symmetry is
\begin{align*}
    \mathrm{Stab}_{\mathcal{G}^{\mathbb{C}}}(\mathrm{diag} (\alpha_1 , ..., \alpha_N )) = \{ \mathrm{diag}(t_1 , ..., t_N ) : t_i \in \mathbb{C}^{*} \} = (\mathbb{C}^{*})^{\times N}
\end{align*}
We use it to make $w_i = v_i$. Let $p_i$ be diagonal entries of $E(z = 0)$.
\begin{statement}{Problem 3}\normalfont\label{prob3}
 In the chosen gauge solve $\mu_{\infty}^{\mathbb{C}} = 0$.
\end{statement}
\begin{exer}\normalfont
Show that the Hamiltonian of the system is the one for elliptic CM:
\begin{align*}
     H = H_{2}^{eCM}
\end{align*}
\end{exer}
\subsection{Spectral curve}
Analogously to symplectic reduction in rational case one can also compute the $A-$matrix providing the Lax representation for eCM with Lax matrix $(\ref{elax})$. This gives the higher integrals of motions $H_k = \frac{1}{k} \mathrm{Tr} (\underset{z = 0}{\mathrm{Res}} L)^{k}$. Fixing a values $(b_1, ..., b_N ) \in \textit{base space }\mathcal{B} \simeq \mathbb{C}^N$ of $(H_1, ..., H_N )$ we set uniquely the coefficients (that are meromorphic functions of $z$-variable) of the polynomial
\begin{align*}
    P_{b}(\lambda) = \mathrm{det} (\lambda - L(z)).
\end{align*}
The set of $(\lambda , z)$ satisfying the equation $P_{b}(\lambda) = 0$ form a curve $\mathcal{C}_{b} \subset T^{*}C_{\tau}$ of genus $g = N-1$. The projection
\begin{align*}
    \pi: \mathcal{C}_{b} &\rightarrow C_{\tau}\\
    (\lambda, z) &\mapsto z
\end{align*}
is a $N$-sheeted covering. Eigenspaces of $L(z)$ define a line bundle $\mathcal{L} \rightarrow \mathcal{C}_{b}$.
In fact the (complex structure of) $\mathrm{rk} = N$ vector bundle, defined by $\bar{A}$, can be reconstructed from the triple $(\mathcal{C}_{b}, \pi,\mathcal{L})$ as $\pi_{*}\mathcal{L}$. Thus the fiber of $\mathcal{P}_{s}^{\mathbb{C}} \rightarrow \mathcal{B}$ is (an open set of) $\mathrm{Jac}(\mathcal{C}_{b})$. This is an abelian variety so one can regard it as the complex analogue of Liouville torus.
\section{Gauge theories and measures on partitions}\label{gauge-part}
Supersymmetric gauge theories in four dimensions, and five and six dimensional theories compactified to four dimensions, produce a class of complexified statistical models. The random variables of these models are collections of Young diagrams, also known as colored partitions \cite{Nekrasov:2002qd, Nekrasov2015BPSCFTCN}. 
More specifically, this is a result of a two-step localization procedure. First of all, it is well-known since the interpretation of M.~Atiyah and L.~Jeffrey of E.~Witten's topological quantum field theory interpretation \cite{Witten:1988ze} of Donaldson theory \cite{Donaldson:1990kn}, path integral of topologically twisted ${\CalN}=2$ supersymmetric gauge theory, for specific correlation functions, can be interpreted as an integral of equivariant differential forms (albeit in an infinite-dimensional setting). 
This interpretation has been generalized to theories with matter hypermultiplets  \cite{Vafa:1994tf,Labastida:1996tz,Moore:1997pc,Losev:1997tp}. Whenever such interpretation holds, path integral localizes to a finite-dimensional integral over the moduli space of instantons, i.e. solutions to the $F^{+}_{A}=0$ equations modulo suitable gauge group.  This localization is an infinite-dimensional version of the equivalence of different representatives of Euler class of a vector bundle: Poincare dual to the zero locus of generic transverse section,
and the curvature Gauss-Bonnet type form. The next step undertaken in \cite{Nekrasov:2002qd} is to deform the theory in such a way that the path integral measure becomes the equivariant differential form with respect to both the local gauge symmetries and the global symmetries consisting of the rotations of ${\BR}^{4}$ (more generally, isometries of the four-manifold on which one studies the theory), the rotations of gauge frame and possibly rotations of the spaces of multiplicities of matter fields. The question of framing arises when one is working on ${\BR}^{4}$ or other non-compact manifolds. We need to impose the boundary conditions both on gauge fields and gauge transformations one divides by. As a result, the gauge group ${\CalG}_{\infty}$ is a proper normal subgroup in the group ${\CalG}$ of all gauge transformations defined on some compactification of ${\BR}^{4}$, e.g. $S^{4} = {\BR}^{4} \cup {\infty}$. The quotient ${\CalG}/{\CalG}_{\infty}$ is the group of frame rotations. The equivariant parameters $\bf a$ for this group are identified with the vacuum expectation value of the adjoint-valued complex Higgs field, the scalar in the vector multiplet of ${\CalN}=2$ supersymmetry:
\beq
\langle {\phi} \rangle = {\bf a} 
\eeq
The equivariant parameters $({\ve}_{1}, {\ve}_{2})$ for the rotation group $Spin(4)$ are not present in the conventional ${\CalN}=2$ theory. 

The way path integral localizes onto the instanton configurations and then onto the set of special instanton configurations has been described in literature on many occasions, see \cite{Pestun:2016zxk} for a review. 
Using the localization in equivariant cohomology we express the integral as a sum over the corresponding fixed points. The subtlety is that the fixed points do not belong to the actual moduli space of smooth supersymmetric gauge field configurations, such as instantons. Rather, they are parked somewhere on the compactified
moduli space. The points of the compactified moduli space no longer parametrize the solutions to the instanton equations on the smooth four dimensional spacetime. They can be interpreted in algebraic geometry as parametrizing torsion free sheaves on ${\BC\BP}^{2}$ -- 
a compactification of ${\BC}^{2}$, a complex plane, whose underlying four dimensional real space is the original spacetime. Every instanton uniquely corresponds to a locally free, hence torsion-free, sheaf. However, there are more general torsion free sheaves. A limit of a sequence of locally free sheaves could be a torsion free sheaf which is not locally free. Let us explain this in a little more detail.

The ordinary $SU(N)$ instanton gauge field, 
\beq
A = A_{\mu} dx^{\mu} \, , \ F = dA + A \wedge A = \frac 12 F_{\mu\nu} dx^{\mu} \wedge dx^{\nu}
\eeq
solving the anti-self-duality equation
\beq
F^{+} = 0\, , \ F_{\mu\nu} + \frac 12 {\ve}_{\mu\nu\kappa\lambda} F^{\kappa\lambda} = 0
\eeq
defines a holomorphic rank $N$ vector bundle $\CalE$ by means of its space of sections. Namely, for each open set $U \subset \mathbb{C}^2$ the space $\Gamma({\CalE}\vert_{U})$ of local holomorphic sections of $\CalE$
is by definition the space of $s \in C^{\infty}(U, \mathbb{C}^N)$ which solve the auxiliary linear problem:
\beq
 \left( {\bar\partial}_{\bar i} + A_{\bar i} \right) s = 0\, , \ {\bar i} = {\bar 1}, {\bar 2}, 
\label{eq:locsec}
\eeq
where we decompose $A$ with respect to the complex structure on ${\BR}^{4} \approx {\BC}^{2}$
\beq
\begin{aligned}
& z^{1} = x^{1} + {\ii} x^{2}\, , \ z^{2} = x^{3} + {\ii} x^{4}\, , \\
& {\zb}^{\bar 1} = x^{1} - {\ii} x^{2} \, , \ {\zb}^{\bar 2} = x^{3} - {\ii} x^{4} \\
\end{aligned}
\eeq
as follows
\beq
A = \sum_{i=1}^{2} A_{i} dz^{i} + A_{\bar i} d{\bar z}^{\bar i}.
\eeq
The space of local holomorphic sections ${\Gamma}({\CalE}\vert_{U})$ is a module over (representation of) the ring ${\CalO}_{U}$ of local holomorphic functions on $U$: if $s$ solves \eqref{eq:locsec}, then so does $f s$, for any $f = f(z^1, z^2) \in {\CalO}_{U}$. Moreover, these spaces behave in a certain natural way under the operations of restriction and intersection of open sets. In order for this collection of modules (it is called a sheaf, also denoted by $\CalE$) to come from the holomorphic vector bundle, it should be locally free, that is
\beq
{\Gamma}({\CalE}\vert_{U}) \approx 
 \begingroup
{\CalO}_{U}^{\oplus N}. 
\endgroup
\eeq
Indeed, in that case we are able to find a fundamental matrix $\Psi = (s_1, ..., s_n)$ of solutions \eqref{eq:locsec} and recover the gauge field by the formula
\beq\label{eq:AfP}
\begingroup
A_{\bar i} = -\bar{\partial}_{\bar i} \Psi \Psi^{-1}.
\endgroup
\eeq
The abovementioned torsion free sheaves obey a weaker condition: 
if $f \in {\CalO}_{U}$ is such, that $f s = 0$ for non-zero $s \in {\Gamma}({\CalE}\vert_{U})$, 
then $f = 0$. However, given the space of local sections of a torsion free sheaf, one cannot, in general, find a $(0,1)$-gauge field $A_{\bar i}dz^{\bar i}$, so that \eqref{eq:locsec} would hold. The issue is that all $\Psi \in {\Gamma}({\CalE}|_{U})^{\oplus N}$ may degenerate at some point, so that
$A_{\bar i}$ defined from \eqref{eq:AfP} would be singular there. A way to address this issue, by keeping the moduli interpretation of the (partly) compactified moduli space of instantons, is to work on a noncommutative ${\BR}^{4}$ \cite{Nekrasov:1998ss}. 

Without getting into the details of the construction, let us just mention that the fixed points of the symmetry used in our equivariant localization are the torsion free sheaves of rank $N$, which split as direct sums
\beq\label{fixedptsh}
{\CalE}= {\CalI}_{{\lambda}^{(1)}} \oplus \ldots {\CalI}_{{\lambda}^{(N)}}
\eeq
of rank one sheaves, invariant under the complex torus ${\BC}^{\times} \times {\BC}^{\times}$
action on ${\BC\BP}^{2} = {\BC}^{2} \cup {\BC\BP}^{1}_{\infty}$, 
\beq
(t_{1}, t_{2}) \cdot (z_{0} : z_{1} : z_{2}) = (z_{0}: t_{1} z_{1} : t_{2}z_{2})
\eeq
which are, in addition, trivial over ${\BC\BP}^{1}_{\infty}$, i.e. ${{\CalI}_{{\lambda}^{(\alpha)}}}\vert_{{\BC\BP}^{1}_{\infty}} \approx {\CalO}_{{\BC\BP}^{1}_{\infty}}$. Such sheaves ${\CalI}_{{\lambda}^{({\alpha})}}$,  ${\alpha}= 1, \ldots, N$,  are the sheaves of holomorphic functions on ${\BC}^{2}$, which form a finite codimension $k_{\alpha}$ ideal in the ring of polynomials ${\BC}[z_{1}, z_{2}]$. The torus invariance means that such an ideal is generated by monomials. It is easy to see
that the generators of the ideal must have the form
\beq
z_{1}^{i-1} z_{2}^{{\lambda}_{i}^{({\alpha})}} \, , \ i = 1, \ldots, {\ell}_{\alpha}
\eeq
where the integers
${\lambda}^{({\alpha})}_{1}, {\lambda}^{({\alpha})}_{2}, \ldots, {\lambda}^{({\alpha})}_{{\ell}_{\alpha}}$ obey the inequality
\beq
{\lambda}^{({\alpha})}_{1} \geq {\lambda}^{({\alpha})}_{2} \geq  \ldots \geq {\lambda}^{({\alpha})}_{{\ell}_{\alpha}}
\eeq
in other words, form a partition. Equivariant localization procedure assigns a weight to each fixed point of the form (\ref{fixedptsh}), and thus defines a measure on a set of multi--partitions. In Sections {\bf 6.2--6.3} we introduce some basic notions relevant for partitions and discuss their properties. Then in Sections {\bf 6.4--6.8} we give precise formulas for the measures on multi--partitions arising from gauge theories in various examples.

To proceed, we need to be more specific about the symmetries used in equivariant localization.
The standard gauge theory on ${\BR}^{4}$ is invariant under the group $Spin(4)$ of Euclidean rotations. The noncommutative deformation of the theory is invariant under the subgroup $U(2) \subset Spin(4)$, preserving the K{\"a}hler form used in deformation quantization of ${\BR}^{4}$. Namely, we deform the commutative coordinate ring generated by $z_1, z_2, {\bar z}_{1}, {\bar z}_{2}$ to the noncommutative algebra with the same generators, with the deformed relations:
\beq
\begin{aligned}
& [z_{1}, z_{2} ] = 0\, , \ [{\bar z}_{1}, {\bar z}_{2}] = 0 \,  ,\\
& [z_{1}, {\bar z}_{1}] = {\theta}_{1}\, , \ [ z_{2}, {\bar z}_{2} ]= {\theta}_{2}\,  , 
\end{aligned}
\label{eq:noncalg}
\eeq
with some real parameters ${\theta}_{1}, {\theta}_{2} \in {\BR}$, 
endowed with the same flat metric 
\beq
dz_{1} d{\bar z}_{1} + dz_{2} d{\bar z}_{2}\ . 
\label{eq:flmt}
\eeq
These notions are reviewed in \cite{Nekrasov:1998ss}.  
Both \eqref{eq:noncalg} and \eqref{eq:flmt} are invariant under the transformations
\beq
(z_{1}, z_{2}) \mapsto  \left( a z_{1} + b z_{2} \, , \ - {\bar b} z_{1} + {\bar a} z_{2} \right)
\eeq
with $|a|^2+ |b|^2 = 1$ and 
\beq
(z_{1}, z_{2}) \mapsto (c z_{1}, c z_{2}) 
\eeq
with $|c| = 1$. These transformations form the $U(2)$ subgroup of $Spin(4)$. When ${\theta}_{1} = {\theta}_{2} = 0$ the symmetry of all $SO(4)$ rotations of $(z_1, z_2, {\bar z}_1, {\bar z}_2)$
is restored. 
\subsection{String theory perspective}
Noncommutative gauge theories \cite{Douglas:2001ba} require string theory for their definition. 
In string theory realization of gauge theory, typically, there is an infinite number of degrees of freedom involved, of which only a finite number is relevant for low energy physics. 
Specifically, the gauge origami \cite{Nekrasov:2016ydq} setup embeds the four dimensional 
spacetime ${\BC}^{2} \approx {\BR}^{4}$ as a subspace of ${\BC}^{4} \approx {\BR}^{8}$. The ten dimensional space-time of Type IIB string theory used in many of these constructions, is ${\BR}^{8} \times {\BT}^{2}$. Metrically, it is the twisted product, i.e.
the total space of rank $4$ vector bundle over ${\BT}^{2}$ with the $SU(4)$ structure group. 
As a complex manifold it is the total space of direct sum of four line bundles
$L_1, L_2, L_3, L_4$ over an elliptic curve $E$, such that the tensor product
$L_1 \otimes L_2 \otimes L_3 \otimes L_4 \approx {\CalO}_{E}$. The parameters $L_1, L_2, L_3, L_4 \in Pic_{0}(E)$ of this background
${\ve}_{1}, {\ve}_{2}, {\ve}_{3}, {\ve}_{4}$ are usually viewed additively, so
\beq
{\ve}_{1} + {\ve}_{2} + {\ve}_{3} + {\ve}_{4} = 0
\eeq
Metrically, the ten dimensional space is flat, with the metric
\beq
ds^{2}_{10} = \frac{A}{\tau_2} dzd{\bar z} + \sum_{a=1}^{4} \left( dZ^a + \frac{\ii\pi}{\tau_2} \left( {\ve}_{a} d{\bar z} + {\bar\ve}_{a} dz \right) Z^a \right) \left( d{\bar Z}^a -\frac{\ii\pi}{\tau_2} \left( {\ve}_{a} d{\bar z} + {\bar\ve}_{a} dz \right) {\bar Z}^a   \right)
\label{eq:IIBmet}
\eeq
The metric \eqref{eq:IIBmet} is written in the complex coordinates $(z, Z^1, Z^2, Z^3, Z^4)$
on a fivefold ${\BC}^{5}$. 
Specifically $z, {\bar z}$ are the additive coordinates on $E$, $z \sim z+ m + n {\tau}$, $m,n \in {\BZ}$, ${\tau} = {\tau}_{1} + {\ii} {\tau}_{2}$, ${\tau}_{1} \in {\BR}$, ${\tau}_{2} \in {\BR}_{+}$. 
The metric \eqref{eq:IIBmet} is not hermitian with respect to the complex structure where
$(z, Z^a)$ are holomorphic coordinates. The parameter $A$  is the area of $T^2  \approx E$ measured in the metric \eqref{eq:IIBmet}.

Consider the diffeomorphism
\beq
(z, {\bar z},  Z^a, {\bar Z}^a) \mapsto \left(z, {\bar z}, Z^a e^{\frac{\pi}{\tau_2} \left( m_a (z-{\bar z}) + n_a ({\bar\tau}{ z} - {\tau} {\bar z} ) \right)} , {\bar Z}^{a} e^{-\frac{\pi}{\tau_2} \left( m_a (z-{\bar z}) + n_a ({\bar\tau}{ z} - {\tau} {\bar z} ) \right)} \right)
\label{eq:dehn}
\eeq
with integers $m_1, m_2, m_3, m_4, n_1, n_2, n_3,n_4$ obeying 
\beq
m_1 + m_2 + m_3 + m_4 = 0\, , \ n_1 + n_2 + n_3 + n_4 = 0 
\eeq
Obviously, the pullback of \eqref{eq:IIBmet} with respect to \eqref{eq:dehn} is again \eqref{eq:IIBmet} with the parameters
\beq
{\ve}_{a} \mapsto {\ve}_{a} + m_{a} + {\tau} n_{a}\, , \ a = 1, 2, 3, 4
\label{eq:shifts4}
\eeq
Together with the obvious permutation symmetry, the transformations \eqref{eq:shifts4} form the affine Weyl group $W_{{\hat A}_3}$ of $\mathfrak{sl}_4$. The associated  space ${\BC}^{3}/W_{{\hat A}_3}$ of equivalence classes of $\ve$'s  is the coarse moduli space of principal holomorphic $SL(4)$-bundles over $E$, also isomorphic to the moduli space of flat $SU(4)$
connections on $T^2$.  

Of course, one can get a more general flat ten dimensional metrics, by using a generic flat $SO(8)$
connection on $T^2$. The latter abelianizes to the maximal torus $U(1)^4$ of $SO(8)$. The significance of the reduction to $SU(4)$ with the maximal torus $U(1)^3 \subset U(1)^4$ is the 
preservation of some nilpotent symmetry, supersymmetry. Mathematically this is saying that
we have an infinite-dimensional complex of (string) fields, whose cohomology is of interest to us. 

This is the data of Type IIB closed string theory on a twisted flat ten dimensional background. 
The twist breaks the maximal ${\CalN}=(0,2)$, $d=10$ supersymmetry to a smaller sub-superalgebra, containing at least the ${\CalN}=(0,2)$, $d=2$ supersymmetry, whose bosonic part consists of  the translations along $E$. The notation $(0,2)$ means that the $z$- and ${\bar z}$- translations do not enter the algebra in a symmetric fashion. There are fermionic generators squaring to ${\bar\partial}_{\bar z}$, but not to ${\partial}_{z}$. 

Our construction does not stop here. We need to add a stack of $N$ $D5$ branes wrapping the total space of $L_{1} \oplus L_{2}$ vector bundle over $E$, i.e. the six dimensional submanifold of the ten dimensional ambient space. In other words, the worldvolume of the $D5$ branes is the zero section of $L_3 \oplus L_4$, taken with multiplicity $N$. 

In addition to the abovementioned geometric parameters $A, \tau, {\ve}_{1}, {\ve}_{2}, {\ve}_{3}$, the Type IIB string background depends on the Type IIB axion-dilaton field $\CalT = a + {\ii} e^{-\phi}$, and the periods $B, B'$ of the NSNS and RR $2$-forms
\beq
B = \int_{T^2} B_{NS}\, , \ B' = \int_{T^2} B_{RR}
\eeq
Macroscopically, the theory on the stack of $D5$ branes is four dimensional, and at low energy
it should be described by a deformed version of ${\CalN}=4$ super-Yang--Mills theory with 
gauge group $U(N)$. The complexified coupling of this gauge theory is 
\beq
e^{-\phi} A + {\ii} \left( a B + e^{-\phi} B' \right)  
\eeq 
One can study various limits where the torus $T^2$ shrinks. It could shrink to $S^1$ or to a point. One scales the dilaton field $\phi$ to keep the effective gauge coupling in five or four dimensions
finite. Mathematically, the theory on finite $T^2$ corresponds to elliptic cohomology of the moduli space of instantons on ${\bR}^{4}$, the theory on $T^2$ collapsed to $S^1$ corresponds to $K$-theory, and the theory on $T^2$ collapsed to a point is the ordinary cohomology theory. 
Some details of this construction can be found in \cite{Nekrasov:1996cz,Baulieu:1997nj}.

\subsection{Partitions}
\begin{defn}\normalfont
    A \textit{partition} is a set of non-increasing integer numbers:
\begin{align*}
    \lambda = (\lambda_1 \geq \lambda_2 \ge  ... \ge \lambda_{l(\lambda)} > 0).
\end{align*}
The $l(\lambda)$ is called the length of $\lambda$. The size of $\lambda$ is the sum $|\lambda| = \lambda_1 + ... + \lambda_{l(\lambda)}$. We denote by $\Pi$ the set of all partitions.
\end{defn}
It is convenient to represent a partition with a Young diagram, i.e. with a subset
\begin{align*}
    \{ \text{box } \square = (i, j) \in \mathbb{Z} : 1 \leq j \leq \lambda_i \}.
\end{align*}
of $\mathbb{Z}^{2}$. The map $(i, j) \mapsto (j, i)$ defines an involution $\lambda \mapsto \lambda^{t}$ on the $\Pi$. 
\begin{nt}\normalfont
Let $\lambda$ be a partition. The $\textit{inner boundary}$ of $\lambda$ is
\begin{align*}
    \partial_{-}\lambda = \{(i, j) \in \lambda: (i+1, j), (i, j+1) \notin \lambda\},
\end{align*}
the \emph{outer boundary} of $\lambda$ is
\begin{align*}
    \partial_{+}\lambda = \{(1, \lambda_1 + 1), (l(\lambda) + 1, 1)\} \cup \{(i, j) \notin \lambda : (i-1, j), (i, j-1) \in \lambda\}.
\end{align*}
\begin{figure}[h]
\centering
\includegraphics[scale=0.5]{partition.png}
\caption{A Young diagram with boxes in $\partial_{+}\lambda$ and $\partial_{-}\lambda$ colored green and pink respectively.}
\end{figure}
\begin{rem}\normalfont
   Note that $\#\partial_{+}\lambda - \#\partial_{-}\lambda = 1$. 
\end{rem}
For the box $\square = (i, j) \in \lambda$ the arm length of $(i, j)$ is
    \begin{align*}
        &a_{\square} = \lambda_i - j,
    \end{align*}
the leg length of $(i, j)$ is
    \begin{align*}
    &l_{\square} = \lambda_{j}^{t} - i
    \end{align*}

\begin{figure}[h]

\begin{subfigure}{0.5\textwidth}
\includegraphics[scale=0.5]{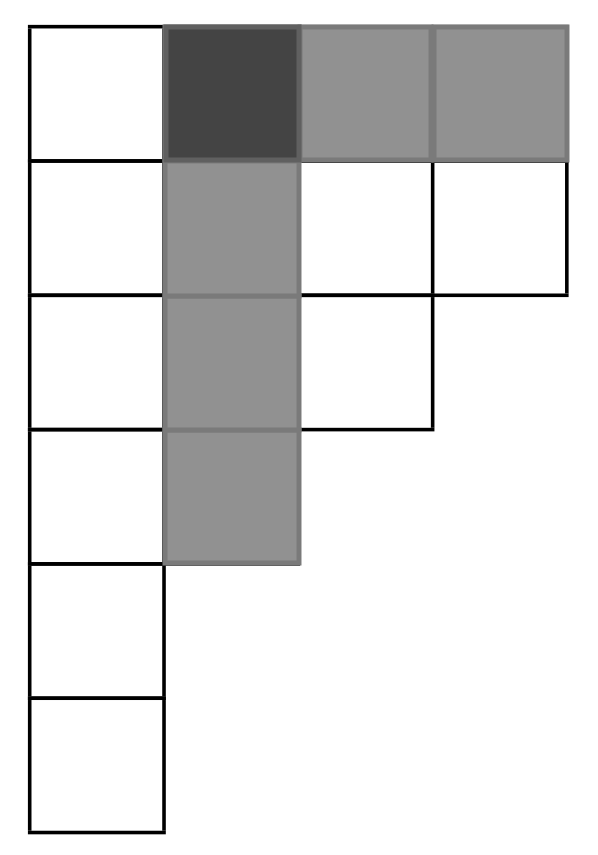} 
\caption{For the box $\square= (1, 2)$ in this diagram, \\we have $a_{\square} = 2$ and $l_{\square} = 3$.}
\end{subfigure}
\begin{subfigure}{0.5\textwidth}
\includegraphics[scale=0.5]{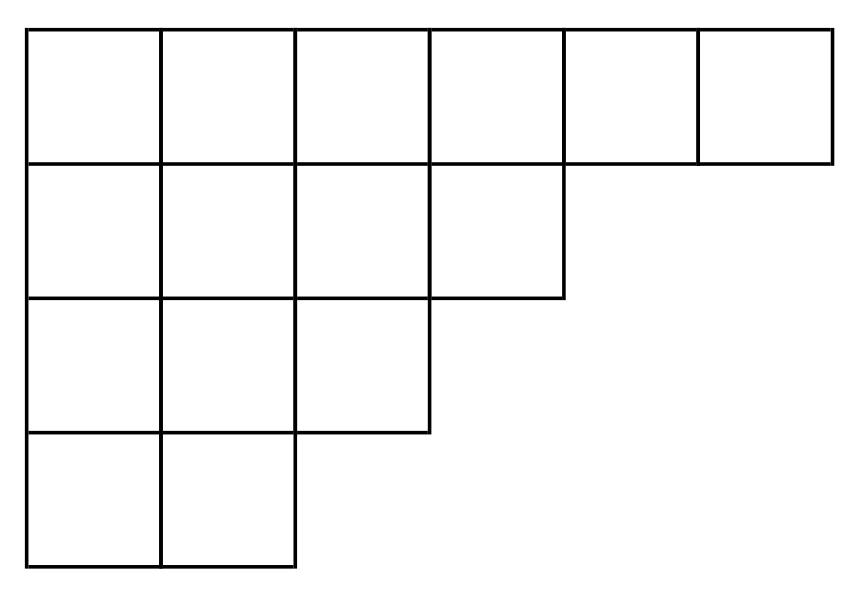}
\caption{This diagram is transposed to the one on the left.}
\end{subfigure}
\caption{A pair of transposed Young diagrams}
\end{figure}

\end{nt}

\begin{rem}\normalfont
   There is a bijection
\begin{align*}
    \{\lambda \in \Pi : |\lambda| = n \} &\leftrightarrow \{\text{monomial ideals in } \mathbb{C}[z_1 , z_2 ] \text{ of codimension } n\}\\
    \lambda & \mapsto J_{\lambda} = \big( \{ z_1^{i-1} z_2^{\lambda_i} \}_{i = 1, ..., l(\lambda)} \big) = \big( \{ z_1^{a-1} z_2^{b-1} \}_{(a, b) \in \partial_{+}\lambda} \big).
\end{align*}
\end{rem}

\begin{exer}\normalfont
Let us define a vector space 
\begin{align}\label{klam}
    K_{\lambda} = \mathbb{C}[z_1 , z_2]/J_{\lambda}.
\end{align}    
Check that $\dim K_{\lambda} = |\lambda|$ and prove the exactness of the sequence
\begin{align}\label{resj}
    0 \rightarrow \bigoplus\limits_{(a, b) \in \partial_{-}\lambda} \mathbb{C}[z_1 , z_2] z_1^{a} z_2^{b} \rightarrow \bigoplus\limits_{(a, b) \in \partial_{+}\lambda} \mathbb{C}[z_1 , z_2] z_1^{a-1} z_2^{b-1} \rightarrow J_{\lambda} \rightarrow 0
\end{align}
\end{exer}
\begin{ex}
    For $\lambda = (5, 3, 2, 2)$ $K_{\lambda}$ has a basis given by the following monomials
\begin{align*}
    \begin{matrix}
    1     &\;\; z_2       &\;\; z_2^2     & \;\; z_2^3 &\;\; z_2^4\\\\
    z_1   &\;\; z_1 z_2   &\;\; z_1 z_2^2 & & \\\\
    z_1^2 &\;\; z_1^2 z_2 &\;\;           & & \\\\
    z_1^3 &\;\; z_1^3 z_2 &\;\;           & &
    \end{matrix}
\end{align*}
\end{ex}
\begin{rem}\normalfont
    Note that the monomial ideals are exactly those which are invariant under the action $(z_1 , z_2) \mapsto (t_1 z_1 , t_2 z_2)$ of $\mathbb{C}^{*} \times \mathbb{C}^{*} \ni (t_1, t_2)$ on $\mathbb{C}^{2}$.
\end{rem}
\subsection{Characters}
\begin{defn}\normalfont
The \textit{character} of a partition $\lambda$ is a polynomial
    \begin{align}\label{chlam}
    \mathrm{ch} \lambda = \chi _{\lambda} (q_1 , q_2 ) = \sum\limits_{\square \in \lambda} q_1^{i-1} q_2^{j-1} \in \mathbb{C}[q_1 , q_2].
\end{align}
Another common notation for such monomials is $q_1^{i-1}q_{2}^{j-1}=e^{c(\square = (i, j))}$.
Let ${\sf R} = {\BC}[[q_{1}, q_{2}]][q_{1}^{-1}, q_{2}^{-1}]$. 
\end{defn}
\begin{defn}\normalfont
    Let $M$ be a $\mathbb{C}^{*} \times \mathbb{C}^{*}$-module. It is just a $\mathbb{C}$-vector space equipped with a linear action of $\mathbb{C}^{*}\times\mathbb{C}^{*} \ni (q_1, q_2)$. It has a decomposition $M = \bigoplus\limits_{(i,j)\in \mathbb{Z}^{2}} M^{(i,j)}$, where $M^{(i, j)} = \{m \in M:  (q_1, q_2).m = q_{1}^{i-1}q_{2}^{j-1}m\}$. Let $M^{(i, j)}=0$ for $i \ll 0$ or $j\ll0$. The \textit{character} of $M$ is a formal power series
    \begin{align*}
    \mathrm{ch} M = \sum\limits_{(i, j) \in \mathbb{Z}^2} \dim M^{(i, j)} q_1^{i-1} q_2^{j-1} \in {\sf R} 
\end{align*}
\end{defn}
\begin{rem}\normalfont
    The formula (\ref{chlam}) is nothing else but the character of $K_{\lambda}$.
\end{rem}
For any bigraded space $M$ and $(m, n) \in \mathbb{Z}^{2}$ one can introduce bigraded space $M_{[m, n]}$ with shifted grading:
\begin{align*}
    M_{[m, n]}^{(i, j)} = M^{(i+m, j+n)}.
\end{align*}
\begin{defn}\normalfont
    Let $M_{\bullet}$ be a bounded chain complex of bigraded vector spaces. The \textit{character} of $M_{\bullet}$ is a formal power series
    \begin{align}\label{chd}
    \mathrm{ch} ({M}_{\bullet}) = \sum\limits_{k \in \mathbb{Z}} (-1)^{k} \mathrm{ch} ( {M}_{k} ) \in {\sf R} .
\end{align}
\end{defn}
\begin{rem}\normalfont
Let $d_{k}: M_{k} \rightarrow M_{k+1}$ be the differentials of $M_{\bullet}$. We have exact sequences
\begin{align*}
    &0 \rightarrow \mathrm{Ker} d_{k} \rightarrow M_{k} \rightarrow \mathrm{Im} d_k \rightarrow 0\\
    &0 \rightarrow \mathrm{Im} d_{k+1} \rightarrow \mathrm{Ker} d_{k} \rightarrow H_{k}(M_{\bullet}) \rightarrow 0,
\end{align*}
which imply
\begin{align*}
    &\mathrm{ch}(M_{k}) = \mathrm{ch}(\mathrm{Ker} d_{k}) + \mathrm{ch}(\mathrm{Im}d_{k})\\
    &\mathrm{ch}(H_{k}(M_{\bullet})) = \mathrm{ch}(\mathrm{Ker} d_{k}) - \mathrm{ch}(\mathrm{Im}d_{k+1}).
\end{align*}
 Thus, we have
\begin{align}\label{chh}
    \mathrm{ch} ({M}_{\bullet}) &= 
    \sum\limits_{k \in \mathbb{Z}} (-1)^{k} \left( \mathrm{ch}(\mathrm{Ker} d_{k}) + \mathrm{ch}(\mathrm{Im}d_{k}) \right) 
    =\sum\limits_{k \in \mathbb{Z}} (-1)^{k} \left( \mathrm{ch}(\mathrm{Ker} d_{k}) - \mathrm{ch}(\mathrm{Im}d_{k+1}) \right) =\nonumber\\
    &=\sum\limits_{k \in \mathbb{Z}} (-1)^{k} \mathrm{ch}(H_{k}(M_{\bullet})).
\end{align}
\end{rem}
\begin{defn}\normalfont
    The $\textit{tautological complex}$ $C^{\lambda}_{\bullet}$ associated to the partition $\lambda$ is a chain complex
    \begin{align*}
    0 \longrightarrow K_{\lambda} \underset{\mathbb{C}}{\otimes} \wedge^{2} Q \stackrel{d_{2}}{\longrightarrow} K_{\lambda} \underset{\mathbb{C}}{\otimes} Q \oplus \mathbb{C} \stackrel{d_{1}}{\longrightarrow} K_{\lambda}\longrightarrow 0,
\end{align*}
where $\mathbb{C}$ is the trivial $\mathbb{C}^{*} \times \mathbb{C}^{*}$-module and $Q = \mathbb{C} z_2 \oplus \mathbb{C} z_1$.
Using the natural isomorphisms
    $K_{\lambda}  \underset{\mathbb{C}}{\otimes} Q \simeq K_{\lambda} \underset{\mathbb{C}}{\otimes} z_2 \oplus K_{\lambda} \underset{\mathbb{C}}{\otimes} z_1 ,\;
    K_{\lambda} \underset{\mathbb{C}}{\otimes} \wedge^{2} Q \simeq K_{\lambda} \underset{\mathbb{C}}{\otimes} z_1 z_2$
we define differentials by the formulas
\begin{align*}
    d_{2}(k \otimes z_1 z_2) &= z_1 k \otimes z_2 + z_2 k \otimes z_1\\
    d_{1}(k_1 \otimes z_2 + k_2 \otimes z_1 + u) &= z_2 k_1 - z_{1} k_2 + u.
\end{align*}
\end{defn}
\begin{exer}
    Check that $d_1 \circ d_{2} = 0$.
\end{exer}
\begin{exer}
    Check that $H_{0}(C^{\lambda}_{\bullet}) = 0$. Also compute
    \begin{align*}
        \mathrm{ch} (H_{1}(C^{\lambda}_{\bullet})) = \sum\limits_{\square \in \partial_{+} \lambda} e^{c(\square)},\\
        \mathrm{ch} (H_{2}(C^{\lambda}_{\bullet})) = \sum\limits_{\square \in \partial_{-}\lambda} e^{c(\square)}.
    \end{align*}
\end{exer}
Let us consider the quantity
\begin{align*}
    s_{\lambda} = - \mathrm{ch}(C^{\lambda}_{\bullet}).
\end{align*}
From the exercises and $(\ref{chh})$ we find
\begin{align}\label{boundif}
    s_{\lambda} = \sum\limits_{\square \in \partial_{+} \lambda} e^{c(\square)} -\sum\limits_{\square \in \partial_{-}\lambda} e^{c(\square)}
\end{align}
On the other hand by the definition $(\ref{chd})$:
\begin{align}\label{sdef}
    s_{\lambda} = 1 - (1 - q_1 )(1 - q_2 )\chi _{\lambda}(q_1 , q_2 ) 
\end{align}
\begin{exer}\normalfont
    Compute
    \begin{align}
        \mathrm{ch} ( \mathbb{C}[q_1 , q_2 ] ) = \frac{1}{(1 - q_1 )(1 - q_2 )}.
    \end{align}
Using the exact sequences $(\ref{klam})$ and $(\ref{resj})$ check one more way the coincidence of $(\ref{boundif})$ and $(\ref{sdef})$.
\end{exer}
These formulas are useful for computing Euler characteristics in $\mathbb{C}^{*}\times\mathbb{C}^{*}$-equivariant $K$-theory. In the following we define some measures that are obtained as a result of such computations.
\subsection{Measures}
Four dimensional gauge theories have special solutions of equations of motions (critical points of the action)  called instantons (\cite{Nekrasov:2005wp}, \cite{FRENKEL2007215}). 
As we explained in the introduction to this section, equivariant
localization of path integrals (\cite{Nekrasov:2002qd}) done in two steps: first to instanton moduli spaces, then to fixed points of global symmetry acting on the moduli space produces  interesting measures $\mu = \mu_{(\lambda^{(\alpha)})}$ on the set of $\textit{multi--partitions}$
\begin{align*}
    \{(\lambda^{\alpha})_{\alpha \in A}: \lambda^{\alpha} \in \Pi \}.
\end{align*}
Here $A$ is some index set. The measures depend on additional complex parameters. 
Typically, the space parameters can be decomposed as the product of the space ${\BC}^{A}$ of \emph{Coulomb moduli and fundamental masses} $a_{\alpha}$, ${\alpha} \in A$,  the space ${\BC}^{E}$ of \emph{bi-fundamental masses}, $m_{e}$, $e \in E$, the space ${\BC}^{3}$ of parameters 
$({\ve}_{1}, {\ve}_{2}, {\ve}_{3})$ of $\Omega$-background, and the space ${\BC}^{\CalV}_{|{\cdot}|<1}$ of \emph{instanton fugacities} ${\qe}_{v}$, $v \in {\CalV}$, or \emph{gauge couplings}
${\tau}_{v} = \frac{1}{2\pi\ii} {\log}({\qe}_{v})$. The set $A$ of Coulomb moduli maps to the set $\CalV$ of gauge couplings by the map 
\beq
{\nu} : A \to {\CalV}
\eeq
The corresponding $\textit{partition function}$
\begin{align*}
    Z^{inst} = \sum\limits_{(\lambda^{(\alpha)})} \mu_{(\lambda^{(\alpha)})} ( {\bf a}, {\bf m},  {\bve} ) \, \prod_{\alpha} {\qe}_{{\nu}({\alpha})}^{|{\lambda}^{({\alpha})}|}
\end{align*}
is a function of these parameters. 
There is a class of models, labelled by a choice of a \emph{quiver}, i.e. an oriented graph 
$\Gamma$. The set $V$ is the set of vertices, the set $E$ is the set of edges. There are two maps
$s,t: E \to V$, sending the edge to its source and target, respectively. 

Define the function $n: {\CalV} \to {\BZ}_{+}$  by
\beq
n_{v} = \# \{ {\nu}^{-1}(v) \} 
\eeq
The nodes $w \in {\CalV}$, such that ${\qe}_{w} = 0$ are called the \emph{frozen}, or \emph{flavor} nodes. Let $W \subset {\CalV}$ denote the set of frozen nodes. By definition, ${\lambda}^{({\alpha})} = {\emptyset}$ for $\alpha \in {\nu}^{-1}(W)$. 
Let us denote by
\beq
V = {\CalV} \backslash W
\eeq
the set of \emph{gauge nodes}. By definition, $0 < |{\qe}_{v}| < 1$, for $v \in V$.

From now on, we call the parameters $a_{\alpha}$, ${\nu}({\alpha}) \in W$ the \emph{fundamental masses}. We call the parameters $a_{\alpha}$, ${\nu}({\alpha}) \in V$ the \emph{Coulomb moduli}.

For a frozen node $w \in W$, denote by $S_{w} = \left( s (t^{-1}(w)) \cup t(s^{-1}(w)) \right) \cap V$ the set of its gauge neighbors. Define, for $v \in V$, 
\beq
m_{v} = \sum_{w \in W\, , \ v \in S_{w}} n_{v} \geq 0
\eeq
the \emph{number of flavors, charged under the node $v$}.

\subsection{Examples based on Dynkin diagrams}
When the graph $\Gamma$ corresponds to an (affine) simply-laced Dynkin diagram we call the associated model \emph{the $\Gamma$-theory}.

For example, the $\widehat{{A}_{r}}$-theory has
$W = {\emptyset}$, 
\beq
V = \{ {\bullet}_{0}, {\bullet}_{1},  \ldots , {\bullet}_{r} \} \, , \ E = \{ _{0}{\rightarrow}_{1}\, , \, _{1}{\rightarrow}_{2}\, , \, \ldots \, , \, _{r-1}{\rightarrow}_{0} \} \, , 
\label{eq:athve}
\eeq
with the maps
\beq
s( _{i}{\rightarrow}_{i+1} ) = {\bullet}_{i} \, , \ t( _{i}{\rightarrow}_{i+1} )  = {\bullet}_{i+1}
\, , \ r \sim 0
\eeq
Let ${\bf N} = \{ 1, \ldots, N \}$. For reasons explained in \cite{Nekrasov2015BPSCFTCN} the set $A$ is the direct product $A = V \times {\bf N}$, with 
the map ${\nu}$ being the projection onto the first factor. 

As another example, the ${\hat D}_{r}$-theory has $W = {\emptyset}$, 
\beq
V = \{ {\bullet}_{0}, {\bullet}_{1},  \ldots , {\bullet}_{r} \} \, , \ E = \{ _{0}{\rightarrow}_{2}\, , \, _{1}{\rightarrow}_{2}\, , _{2}{\rightarrow}_{3}\, \, 
\ldots \, , _{r-3}{\rightarrow}_{r-2} \, , \, _{r-2}{\rightarrow}_{r}\, , \ _{r-2}{\rightarrow}_{r-1} \} \, , 
\label{eq:dthve}
\eeq
with the maps
\beq
s( _{i}{\rightarrow}_{j} ) = {\bullet}_{i} \, , \ t( _{i}{\rightarrow}_{j} )  = {\bullet}_{j}
\, , \ 
\eeq
Here, the set $E$ contains $_{i}{\rightarrow}_{j}$ where either $j=i+1$ for $i=1, \ldots, r-2$, or $i=0$ and $j=2$, or
 $i=r-2$ and $j=r$. Let ${\bf N}_2 = \{ 1, \ldots, 2N \} \approx {\bf N} \amalg {\bf N}$. For reasons explained in \cite{Nekrasov2015BPSCFTCN} the set $A$ is the union of the
direct products $A = \{ 0, 1, r-1, r \} \times {\bf N} \amalg \{ 2, \ldots, r-2 \} \times {\bf N}_{2}$, with 
the map ${\nu}$ being the projection onto the first factor in each summand. 

As a last example in this section, we present the $A_r$-theory. Unless additionally specified, 
$W = \{ 0, r+1 \}$, $V = \{ 1, \ldots, r \}$, $E = \{ _{0}{\rightarrow}_{1}, \, \ldots \, , \, _{r}{\rightarrow}_{r+1}\}$, and the maps $s(_{i}{\rightarrow}_{i+1}) = i$, $t(_{i}{\rightarrow}_{i+1}) = i+1$, for $i = 0, 1, \ldots, r$. 

\subsection{Three theories}

In agreement with the theory \cite{Arnold:1997} of mathematical trinities
there are three versions of everything worth discussing. In our case we have the cohomological $H$, $K$-theoretic $K$, and elliptic cohomology $Ell$ versions of the model. These models originate in four-dimensional gauge theory, five-dimensional gauge theory compactified on a circle, and six-dimensional gauge theory compactified on a two-torus. 

In the latter case the measure depends, in addition to everything else, on the modulus ${\sigma}$, ${\Im}{\sigma} > 0$ of the elliptic curve
\begin{align*}
    \mathfrak{p} = e^{2 \pi {\ii} \sigma} \ .
\end{align*}
Depending on the  cohomology theory under consideration,  
the measure is expressed in terms  of the function $\theta(x)$, defined by:
\begin{align*}
\theta(x) = 
    \begin{cases}
        x, & \;H\, {\rm or}\, 4d\\
        1 - e^{-x}, & \;K\, {\rm or} \, 5d\\
        \prod\limits_{n = 1}^{\infty} (1 - \frak{p}^{n})(1 - \frak{p}^{n-1} e^{-x})(1 - \frak{p}^{n} e^{x}), & \;Ell\, {\rm or} \, 6d.
    \end{cases}
\end{align*}

\subsection{Example: $\widehat{A_{0}}$-theory}
Let us discuss the $\widehat{A_{0}}$-theory in detail. It is the quiver model in which both 
sets $V = \{ {\bullet} \}$ and $E = \{ {\rightarrow} \}$ consist of one element, with the maps
$s({\rightarrow}) = t({\rightarrow}) = {\bullet}$.

    The index set $A$ in that case is just $\{1, ..., N\}$. Its elements are called $\textit{colors}$. The parameters of the measure are
\begin{align*}
    & {\bf a} = (a_1 , ..., a_N ) \in {\BC}^{N}, ~ \sum\limits_{\alpha = 1}^{N}a_{\alpha} = 0\\
    &{\bve} = (\varepsilon_{1}, \varepsilon_{2}, \varepsilon_{3}) \in \mathbb{C}^{3}, ~ {\ve}_{4} = - \sum\limits_{a = 1}^{3} \varepsilon_{a}\\
    &q_a = e^{\varepsilon_a}\\
    &\frak{q}_{\bullet} = {\qe} = e^{2 \pi {\ii} \tau}, ~ {\rm Im}\tau > 0.
\end{align*}

The measure $\mu$ is given by the formula
\begin{align}\label{A0hatm}
    \mu_{(\lambda^{(1)}, ..., \lambda^{(N)})}({\bf a},{\bve} ) &=  \nonumber\\
    &\prod\limits_{\alpha , \beta = 1}^{N}  \prod\limits_{(i, j) \in \lambda^{(\beta)}} \frac{\theta (\varepsilon_3 + \varepsilon_1 (i - \lambda_{j}^{(\beta)t}) + \varepsilon_2 (1 + \lambda_{i}^{(\alpha)} - j) + a_{\alpha} - a_{\beta})}{\theta(\varepsilon_1 (i - \lambda_{j}^{(\beta)t}) + \varepsilon_2 (1 + \lambda_{i}^{(\alpha)} - j) + a_{\alpha} - a_{\beta})}\nonumber\\
   &  \prod\limits_{\alpha , \beta = 1}^{N}  \prod\limits_{(i, j) \in \lambda^{(\alpha)}} \frac{\theta (\varepsilon_3 + \varepsilon_1 (\lambda_{j}^{(\alpha)t} + 1 - i) + \varepsilon_2 (j -\lambda_{i}^{(\beta)}) + a_{\alpha} - a_{\beta})}{\theta(\varepsilon_1 (\lambda_{j}^{(\alpha)t} + 1 - i) + \varepsilon_2 (j -\lambda_{i}^{(\beta)}) + a_{\alpha} - a_{\beta})}
\end{align}
while the instanton factor is given by
\beq
\prod\limits_{\alpha = 1}^{N} \frak{q}^{|\lambda^{(\alpha)}|}
\eeq

\begin{statement}{Problem 4}\normalfont \label{prob4}
For rational case compute the first and second order terms in expansion
\begin{align*}
Z^{inst} = 1 + \mathfrak{q}Z^{inst}_{1} + \mathfrak{q}^2 Z^{inst}_{2} + ...
\end{align*}
\end{statement}
\begin{statement}{Problem 5}\normalfont \label{prob5}
In the limit $\varepsilon_1 , \varepsilon_2 \rightarrow 0$ find the leading asymptotics of $\mathcal{F}^{inst}_{1}$ and check that $\mathcal{F}^{inst}_{2}$ is regular where the instanton prepotential $\mathcal{F}^{inst}$ is defined to be
\begin{align*}
\mathcal{F}^{inst} = \varepsilon_1 \varepsilon_2 \log Z^{inst} = \mathfrak{q} \mathcal{F}^{inst}_{1} + \mathfrak{q}^2 \mathcal{F}^{inst}_{2} + ...
\end{align*}
\end{statement}
\begin{ex}\normalfont
In the case $N=1$ the abovementioned partly compactified moduli space of instantons is the Hilbert scheme of points on ${\BC}^{2}$. The following theorem illustrates the connection of its enumerative geometry to the measures above.
\begin{thm}
    The $N = 1$ partition function has the representation
\begin{align*}
     Z^{N=1} (\tau, {\ve}_{3};\varepsilon_1 , \varepsilon_2 ) = \sum\limits_{n = 0}^{\infty} \frak{q}^{n} \int\limits_{\mathrm{Hilb}^{[n]}(\mathbb{C}^{2})}\sum\limits_{i} (-q_3)^{i} \wedge^{i} T^{*} \mathrm{Hilb}^{[n]}(\mathbb{C}^{2}),
\end{align*}
where $\int\limits_{\mathrm{Hilb}^{[n]}(\mathbb{C}^{2})}$ is a pushforward to the point in appropriate equivariant cohomology theory (\cite{OkoPan2004}).
\end{thm}
\begin{rem}
    In the 4d case with $\theta(x) = x$ this formula reduces \cite{Nekrasov2015BPSCFTCN} to
\begin{align*}
    Z^{N=1} = \big(\prod\limits_{n = 1}^{\infty} (1 - \frak{q}^n ) \big)^{-\frac{({\varepsilon}_3 + {\varepsilon}_1 )({\varepsilon}_3 + {\varepsilon}_2 )}{{\varepsilon}_1 {\varepsilon}_2}}
\end{align*}
\end{rem}
\end{ex}
For $N > 1$ the situation is more complicated. 
\begin{ex}\normalfont
    $N = 2$ partition function $Z^{N=2}$ for cohomological case describes a conformal block for Liouville CFT ($+$ free boson) (\cite{Alday:2009aq}) on pointed elliptic curve $C_{\tau}$.
\end{ex}
\subsection{Vershik-Kerov limit}
Let us take the limit
$\varepsilon_3 \rightarrow \infty , \frak{q} \to 0$, ${\ve}_{1} + {\ve}_{2} \to 0$ of $\frak{q}^{|\lambda|}\mu_{(\lambda)}(\bve)$ while keeping the quantities $\Lambda^2 = {\ve}_3^{2} \frak{q}$
and $\varepsilon_{1} = -\varepsilon_2 = {\hbar}$ finite. The limiting measure is
\beq
\mu_{\lambda} = \frac{\left( - \frac{\Lambda^2}{\hbar^2} \right)^{|{\lambda}|}}{|\lambda|!} \times
\eeq
a normalized inverse product of squares of hook-lengths: 
\begin{align*}
    |\lambda|! \,  \prod\limits_{\square \in \lambda}  \frac{1}{(a_{\square} + l_{\square} + 1)^2} = \frac{(\dim \lambda)^2}{|\lambda|!}.
\end{align*}
The latter is known  as the $\textit{Plancherel}$ measure on the set of representations of a symmetric group.

Now take the limit $\hbar \rightarrow 0$ with finite $\Lambda$. In this limit one finds the study of the limit shape of very large Young diagrams due to A. M. Vershik and S. V. Kerov (\cite{Logan1977AVP}, \cite{VerKer77}, \cite{VerKer85}).

\section{Order and Disorder}

The view on quantum field theory as a limit of statistical mechanics suggests two types of interesting observables. The first type, the so-called order operators, are defined in terms of elementary fields locally, i.e. they are renormalized finite expressions in fields and their derivatives, evaluated at a point of spacetime, on a loop, a surface etc., the main feature being the locality of the associated functional. 
The next class of observables, the disorder operators, are defined in a tricky way. Sometimes they can also be expressed through the elementary fields of the theory, but in a non-local way.  Typically one defines them as a prescription to perform the path integration over the space of fields, which are well-defined outside a submanifold $N \subset M$ of spacetime manifold, 
and have a singularity along $N$ of prescribed type, or asymptotics. In the following we consider examples of these operators appearing in particular quantum field theories. Then we give a hint on how equivariant localization discussed in (\ref{gauge-part}) turns order and disorder observables in four-dimensional supersymmetric gauge theories to specific functions on the set of multi--partitions. 
We do not derive these functions here; instead, we introduce their final expressions as definitions. Then we study their expectation values.
\subsection{Abelian sigma model}
A possible example of a model containing both order and disorder operators
is the two dimensional sigma model valued in $S^1$ or, more generally, a torus $T = V/{\Lambda}$, where $V \approx {\BR}^{n}$ is Euclidean vector space, and $\Lambda \approx {\BZ}^{n}$ is the $n$-dimensional lattice acting on $V$ by translations. 
The theory is specified by a choice of the metric $G$ and antisymmetric bilinear form $B$ on $V$: 
\beq
G : S^{2}V \to {\BR}\, ,\ B : {\wedge}^{2} V \to {\BR}
\eeq
with the help of  the action
\beq
S[ {\varphi} ] = \frac{1}{4\pi} \int_{\Sigma} G \left( d{\varphi} \wedge \star d{\varphi} \right) + B \left( d{\varphi} \wedge d{\varphi} \right)
\eeq
Given $p \in {\Lambda}^{\vee} \subset V^{\vee}$ (the dual lattice is defined as a subset of the dual space $V^{\vee}$ which consists of all covectors taking integer values on $\Lambda \subset V$), we define
\beq
{\CalV}_{p}(x) = : e^{2\pi\ii p\cdot {\varphi}(x)}:
\eeq  
where
${\varphi}(x) \in V$ is a lift of the value of the field at $x$ to $V$ from $T$, and $:  :$ is the sign of the normal ordering. In quantum theory the field $\varphi (x)$ is not well-defined pointwise;  its average
\beq
{\varphi}_{\ep}(x): = \int_{S^{1}_{\ep}(x)} {\varphi}(x') dx' 
\eeq
over a small circle of radius $\ep$ centered at $x$, however, is well defined. So, 
\beq
{\CalV}_{p}(x) = {\rm Lim}_{\ep \downarrow 0} \ e^{2\pi \ii {\varphi}_{\ep}(x)} {\ep}^{\frac{p \cdot G^{-1}p}{2}}\, , 
\label{eq:vertex}
\eeq
 sometimes called the vertex operator, is an order observable in the abelian sigma model.
\subsection{Wilson loop in gauge theory}
Gauge theory also has both order and disorder operators. An example of the order operator is the Wilson loop observable, it is associated to a loop $C$,  an embedded circle, and a representation $R$ of the gauge group:
\beq
W_{R}(C) = {\Tr}_{R} \, P{\exp} \oint_{C} A
\label{eq:wl}
\eeq
In gauge theories with adjoint scalars and fermions one can generalize \eqref{eq:wl} to include those. This would require additional geometric data, though.  For example, people study the Wilson loops of the form
\beq
{\CalW}_{R}(C) = {\Tr}_{R} \, P{\exp} \oint_{C} A + {\ii} {\Phi}
\eeq
where ${\Phi}(t) = \sum_{I=1}^{n} {\phi}^{I} {\dot y}_{I}(t) dt$, where $I$ labels the adjoint scalars in the theory (the ${\CalN}=4$ super-Yang--Mills theory in four dimensions has $n=6$, ${\CalN}=2$ has $n=2$) and ${\vec y}(t)$ is a parametrized
loop in the auxiliary space ${\BR}^{n}$. One can also include adjoint fermions via
\beq
{\tilde\CalW}_{R}(C) = {\Tr}_{R} \, P{\exp} \oint_{C} A + {\vartheta} \cdot {\Psi} + {\ii} {\Phi} + {\vartheta}{\vartheta} F_{A} + \ldots
\eeq
which depend on a loop in ${\BR}^{n|m}$ with some $n,m$ etc. 

\subsection{Monopole operator in 3d Maxwell theory}
The simplest example of such observable is the monopole operator in 3d gauge theory. Let us start with abelian $U(1)$ Maxwell theory in three dimensions
\beq
\int [DA] \, e^{-\frac{1}{4g^2} \int_{M^3} F \wedge \star F}
\label{eq:max3d}
\eeq
where $F = dA$ is the curvature of the $U(1)$ gauge field $A$. In writing \eqref{eq:max3d} we adopted the physics convention where $A$ is real, so that the connection $1$-form is actually ${\ii}A$. Thus, $F$ is a closed $2$-form, with quantized periods:
\beq
\left[ \frac{F}{2\pi} \right] \in H^{2}(M^3, {\BZ})
\label{eq:diracqu}
\eeq
\subsection{Duality between 3d Maxwell theory and $U(1)$ sigma model}
As is well-known, we can equivalently represent the theory \eqref{eq:max3d} by a $U(1)$-valued sigma model. First we write the path integral for a coupled system of unconstrained $2$-form $F$ and $U(1)$-valued scalar field ${\varphi}\sim {\varphi} + 1$:
\beq
\int [DF][D{\varphi}] \, e^{-\frac{1}{4g^2} \int_{M^3} F \wedge \star F} e^{\ii \int F \wedge d{\varphi}}
\label{eq:max3d2}
\eeq
Integrating out $\varphi$, summing over the possible windings $[d\varphi] \in H^{1}(M^3, {\BZ})$ we get the constraints: $dF = 0$, and the cohomology class of $F$ is $2\pi$-integral as in \eqref{eq:diracqu}. On the other hand, integrating out $F$ we get the effective Lagrangian
\beq
\int [D{\varphi}] \, e^{-g^2 \int_{M^3} d{\varphi} \wedge \star d{\varphi}}
\label{eq:max3d3}
\eeq
Now let us ask ourselves, what does the operator
\beq
{\CalO}_{n}(x) = e^{2\pi \ii n {\varphi}(x)}
\label{eq:locmon}
\eeq
represent? Of course, quantum mechanically \eqref{eq:locmon} needs to be regularized (normal-ordered). One way to achieve that is to define
\beq
{\CalO}_{n}(x) = {\rm Lim}_{{\ep} \downarrow 0} \ e^{2\pi \ii n \int_{S^{2}_{\ep}(x)} {\varphi}(x') d^2x'}\ e^{\# \frac{n^2}{g^2 \ep}},
\label{eq:locmon2}
\eeq
where we denoted by $S^{2}_{\ep}(x) \subset M^3$ a $2$-sphere of geodesic radius $\ep \to 0$ centered at $x \in M^3$, a boundary
\beq
S^{2}_{\ep}(x) = {\partial} B^{3}_{\ep}(x) \subset M^3
\eeq
of a small $3$-ball. 
\begin{statement}{Problem 6}\normalfont \label{prob6}
Compute $\#$ in (\ref{eq:locmon2}).
\end{statement}
At any rate, $\CalO_{n}(x)$ is a local operator in the $\varphi$-description. What is it in the $A$-description?
If we repeat the steps \eqref{eq:max3d2}, \eqref{eq:max3d3} with \eqref{eq:locmon2}
inserted, we would get, instead of the constraint $dF = 0$, the constraint
\beq
dF = 2\pi n {\delta}^{(3)}_{x}\,  ,
\eeq
in other words, the $2$-form $F$ has a nontrivial flux $2\pi n$ through an arbitrary
small $2$-sphere $S^2_{\ep}(x)$ surrounding $x \in M^3$. In other words, 
the gauge field $A$ is well defined on $M^3 \backslash B^{3}_{\ep}(x)$, it has a singularity at $x$ modeled on Dirac monopole. The subtraction scheme used in
\eqref{eq:locmon2} ensures the limit $\ep \to 0$ is finite (we roughly subtract 
from the action the infinite energy of Dirac monopole). 

On the other hand, in the original description, the natural order observable was the Wilson loop
\beq\label{abwil}
W_{n}(C) = P e^{\ii n \oint_{C}A}\, , \ n \in {\BZ}\,, 
\eeq
(irreducible representations of $U(1)$ are labelled by integers). 
\begin{statement}{Problem 7}\normalfont \label{prob7}
What does Wilson loop $W_{n}(C)$ correspond to in the $\varphi$ description?
\end{statement}

\subsection{Wilson loop in non-abelian gauge theory}
In non-abelian theory the situation is even more interesting. 

First of all, let us unpack the definition of the Wilson loop. We work with the
compact gauge group $G$. Let us take for simplicity $R$ to be an irreducible representation. 
Recall that $R$ can be viewed as the Hilbert space of an auxiliary quantum mechanical model, whose classical limit, i.e. the phase space, is some coadjoint orbit ${\CalO}_{R} = G/H \subset {\fg}^{*}$ of $G$. The generic coadjoint orbit ${\CalO}_{\xi} \subset {\fg}^{*}$
is the set of covectors, conjugate to $\xi \in {\fg}^{*}$ by the coadjoint group action:
\beq
{\CalO}_{\xi} = \{ \, {\eta}\, | \, {\eta} = Ad_{g}^{*}{\xi}\ {\rm for\ some\ } g \in G \}
\eeq
The coadjoint orbits are classified by the equivalence classes ${\fg}^{*}/G = {\mathfrak{t}}^{*}/W$. 

\begin{ex}
Consider the case $G=SU(N)$. The coadjoint orbits
are isomorphic to adjoint orbits, the latter are classified by the collection $J = \left( {\ii\lambda}_{1}, \ldots , {\ii\lambda}_{N} \right)$ of eigenvalues of the antihermitian $N\times N$ traceless matrix $g^{-1}{\rm diag}(J) g$. Since these are purely imaginary numbers, we can order them: 
\beq
{\lambda}_{1} \geq {\lambda}_{2} \geq \ldots \geq {\lambda}_{N}
\label{eq:eigenvalues}
\eeq
The coadjoint orbit ${\CalO}_{J}$ is the space
of all antihermitian matrices $X$ whose eigenvalues
are given by $J$. Remarkably, this is a complex manifold, isomorphic to the (partial) flag variety
\beq
Fl(v_{1}, v_{2}, \ldots , v_{r}) = GL(N, {\BC})/P_{v_{1}, v_{2}, \ldots , v_{r}}
\label{eq:pflag}
\eeq
with
\beq
N > v_{1} > v_{2} > \ldots > v_{r}
\eeq
where
\beq
m_{1} = N-v_{1}\, , \ m_{2} = v_{1}-v_{2}\, ,  \ldots \, , \ m_{r} = v_{r-1}-v_{r}, m_{r+1} = v_{r}
\eeq
are the multiplicities of the eigenvalues, ordered
as in \eqref{eq:eigenvalues}
\beq
{\lambda}_{1}= \ldots = {\lambda}_{m_{1}} = l_{1} >
{\lambda}_{m_{1}+1}= \ldots = {\lambda}_{m_{1}+m_{2}} = l_{2} > \ldots
> {\lambda}_{N-v_{r}+1} = \ldots = {\lambda}_{N} = l_{r+1}
\eeq
The points of \eqref{eq:pflag} are the flags
$V_{r} \subset V_{r-1} \subset \ldots \subset V_{1} \subset W$
of subspaces of the fixed complex vector space $W = {\BC}^{N}$, with ${\rm dim}V_{i} = v_{i}$. Endowing $W$ with hermitian structure, we identify
\beq
Fl(v_{1}, v_{2}, \ldots , v_{r}) = U(N)/\left( U(m_1)\times \ldots \times U(m_{r+1}) \right)
\eeq
as follows: 
given an operator ${\ii} X \in End(W)$ with hermitian $X = X^{\dagger}$, consider the function
$h_{X}({\psi}) = {\psi}^{\dagger} X {\psi}$
on $W \ni \psi$. It is invariant under the $U(1)$-action ${\psi} \mapsto e^{\ii\alpha}{\psi}$, this, it descends, by restricting it first to $S^{2N-1}  = \{ {\psi}\, | \, {\psi}^{\dagger}{\psi} = 1 \} \subset {\BC}^{N}$, to a function $h_{X}$ on ${\BC\BP}^{N-1} = S^{2N-1}/U(1)$. The critical points of the latter are
the normalized eigenvectors of $X$. The minima correspond to the lowest eigenvalues, the maxima correspond to the maximal eigenvalues. Let $C_{1}, \ldots, C_{r+1}$ be the connected components of the sets of  $h_{X}$ critical points, with
\beq
h_{X}(C_{i}) = l_{i}\, , \ i = 1, \ldots, r+1\ .
\eeq
Let $N_{1}, \ldots, N_{r+1}$ be the descending manifolds for the gradient flow of $h_{X}$ emanating from $C_{1}, \ldots, C_{r+1}$. Their closures are compact submanifolds obeying ${\bar N}_{r+1} \subset {\bar N}_{r} \subset \ldots \subset {\bar N}_{1}$, (obviously ${\bar N}_{1} = \mathbb{CP}^{N-1}$, ${\bar N}_{r+1} = N_{r+1} = C_{r+1}$) the projectivizations of $V_{r} \subset V_{r-1} \subset \ldots \subset V_{1} \subset W$. 
\end{ex}
The coadjoint orbits ${\CalO}_{\xi}$ are the symplectic leaves of the Kirillov--Kostant Poisson structure on ${\fg}^{*}$:
\beq
\{ f, g \}\vert_{\eta} = \left( {\eta} , \left[ df , dg  \right] \right)
\eeq
The corresponding symplectic form ${\omega}_{\xi}$
represents a nontrivial class in $H^{2}({\CalO}_{\xi}, {\BR})$. 
For $G= SU(N)$ the periods of $\omega_{\xi}$ are linear combinations of $l_{i}-l_{i+1}$ for $i = 1, \ldots, r$. 

The coadjoint action of $G$ on ${\CalO}_{\xi}$
preserves ${\omega}_{\xi}$ and has a moment map. In fact, the moment map generating the coadjoint action is the embedding ${\mu}: {\CalO}_{\xi} \rightarrow {\fg}^{*}$. 

When the symplectic form is integral
$[{\omega}_{\xi}] \in H^{2}({\CalO}_{\xi}, {\BZ})$, the associated orbit is called \emph{quantizable}. 
There is a holomorphic line bundle ${\CalL} \to {\CalO}_{\xi}$, such that $c_{1}({\CalL}) = [{\omega}_{\xi}]$. There is a $U(1)$-connection
$\nabla$ on $\CalL$, whose curvature is determined by the symplectic structure
$F_{\nabla} = 2\pi \ii {\omega}_{\xi}$. Geometric quantization associates to $\CalO_{\xi}$ the Hilbert space 
\beq
{\CalH}_{\xi} = H^{0} ( {\CalO}_{\xi} , {\CalL} ) 
\eeq
The group $G_{\BC}$ acts on ${\CalH}_{\xi}$ by geometrically moving the holomorphic sections. The compact group $G$ preserves the inner product. The main result of geometric quantization program is that all the irreducible representations of $G$
are obtained in this way, $R = {\CalH}_{\xi}$, for some ${\xi} = {\xi}_{R}$. 

Let $A = A(t) dt$ be some $\fg$-valued $1$-form
on $S^1$. Consider the monodromy operator $g \in G$ 
\beq
g = P{\exp} \oint A.
\eeq
The Wilson loop observable is given by the character of $g$ in representation $R$:
\begin{align*}
    W_{R} = \chi_{R}(g) = {\Tr}_{R} (g).
\end{align*}
What is important for us
is the path integral representation of it:
\beq
{\chi}_{R}(g) = \int_{{\eta}: S^{1} \to {\CalO}_{\xi}} [D{\eta}(t)] \, e^{2\pi\ii \oint_{S^{1}} \,  d^{-1}{\omega}_{\xi} + \ii \oint_{S^{1}} ({\eta}, A)}.
\label{eq:charpathint}
\eeq
One way to understand \eqref{eq:charpathint} is to use an infinite-dimensional version of Duistermaat-Heckman formula (DH) \cite{DH0,DH1,DH2}. Consider the space $L{\CalO}_{\xi}$ of parametrized loops ${\eta}: S^{1} \to {\CalO}_{\xi}$. It is an infinite-dimensional symplectic manifold with the symplectic form given by the average pullback of ${\omega}_{\xi}$:
\beq
{\Omega}_{L{\CalO}_{\xi}}({\delta}_1{\eta} , {\delta}_{2} {\eta}) = \int_{S^1}\, dt\, {\omega}_{\xi} \left( {\delta}_1 {\eta}(t), {\delta}_2 {\eta}(t) \right).
\label{eq:avsymf}
\eeq
It is invariant under the action of the loop group $LG$ on $L{\CalO}_{\xi}$ by point-wise coadjoint action. It is also invariant under the $U(1)$-action ${\eta} \mapsto {\eta}^{s}$, ${\eta}^{s}(t) = {\eta}(t-s)$. Thus, $\Omega_{L{\CalO}_{\xi}}$
is invariant under the action of the semi-direct
product ${\widetilde{LG}} = U(1) \sdtimes LG$. The moment map for the ${\widetilde{LG}}$ action is not actually globally defined, as the action is not Hamiltonian. It is locally Hamiltonian, though: the one-form on $L{\CalO}_{\xi}$ 
\beq
{\alpha}_{\xi} = \iota_{{\dot\eta}} {\Omega}_{L{\CalO}_{\xi}} = \int_{S^1} dt \, {\omega}_{\xi} ( \dot\eta (t), {\cdot} ) = (p_{2})_{*} {\rm ev}^{*} {\omega}_{\xi}
\eeq
is closed. It is the transgression of the closed two-form ${\omega}_{\xi}$ with respect to the evaluation map
\beq
{\rm ev}: S^1 \times L{\CalO}_{\xi} \rightarrow {\CalO}_{\xi}.
\eeq
Fix a point $p \in {\CalO}_{\xi}$. We can define the function
\beq
{\tilde L}\left({\ell}, [{\sf path}]_{p}^{\ell}\right) = \int_{\sf path} {\alpha}_{\xi} 
\label{eq:ltilde}
\eeq
where ${\sf path}$ is a path on $L{\CalO}_{\xi}$ connecting a given loop ${\ell}$ to the constant loop ${\eta}(t) = p$ (note that the fundamental group of a partial flag variety is trivial for compact connected simply connected $G$). Since the value of the integral does not change under any homotopy preserving the endpoints $\ell$ and $p$, it only depends on the homotopy class $[{\sf path}]_{p}^{\ell}$, also known as the point  
on the universal cover of $L{\CalO}_{\xi}$, covering $\ell \in L{\CalO}_{\xi}$. 

The beauty of quantizable $\CalO_{\xi}$'s is that all the periods of 
${\alpha}_{\xi}$ are integers. Therefore, the function 
\beq
e^{2\pi\ii {\tilde L}}
\eeq
is well-defined on $L{\CalO}_{\xi}$. It is this factor that we denote by
\beq
e^{2\pi\ii \oint_{S^{1}} d^{-1}{\omega}_{\xi}}
\eeq
in \eqref{eq:charpathint}. Now the whole action (integrand in the exponential) in \eqref{eq:charpathint} is nothing but the (multi-valued) moment map for the ${\widetilde{LG}}$ action on $L{\CalO}_{\xi}$, evaluated on the element
\beq
{\phi} = {2\pi\ii} {\partial}_{t} + A(t) \in Lie ({\widetilde{LG}} ) = {\BR} {\partial}_{t} \oplus L {\fg}
\eeq
We are therefore in the realm of the DH formula in the formulation of \cite{DH2}: let ${\BT}$ be the torus of ${\widetilde{LG}}$
generated by $\phi$. By conjugation (Floquet--Lyapunov theory) ${\phi}$ can be brought to the form 
\beq
{\phi} \sim {2\pi\ii}{\partial}_{t} + {\sf a}\, , \ 
\eeq
with constant ${\sf a} \in {\mathfrak{t}} \subset {\fg}$. The DH formula then evaluates \eqref{eq:charpathint} as a sum over the fixed points of ${\tilde g} = e^{{\beta}{\phi}}$ (with generic $\beta$) action on ${L{\CalO}_{\xi}}$. What are these? Since ${\tilde g}$ acts by 
\beq
({\tilde g}\cdot {\eta} ) (t) = Ad_{e^{\beta{\sf a}}}^{*}  {\eta} ( t - 2\pi \beta ) \ .
\eeq
Taking $\beta$-derivative at $\beta =0$ we arrive at the equation
\beq
2{\pi}{\partial}_{t}{\eta} - ad^{*}_{\sf a}({\eta}) = 0
\label{eq:etat}
\eeq
with constant ${\sf a}$. Evaluating \eqref{eq:etat} on ${\mathfrak{t}} \ni {\sf v}$ we get 
that ${\eta}\vert_{{\mathfrak{t}}} = const$. Evaluating \eqref{eq:etat} on a root subspace ${\fg}_{\alpha}$, we get
\beq
{\eta} (t) \vert_{{\fg}_{\alpha}}\ = \ e^{-\frac{{\alpha}({\sf a})}{2\pi} t} {\eta} (0) \vert_{{\fg}_{\alpha}}.
\eeq
Supposing that $g$ (and hence ${\sf a}$) is general, we see that ${\eta} (0) \vert_{{\fg}_{\alpha}}$ must be zero in order to ${\eta} (t) \vert_{{\fg}_{\alpha}}$ be periodic, so
\begin{align*}
{\eta} (t) \vert_{{\fg}_{\alpha}} \ \equiv \ 0.
\end{align*}
Thus, the fixed points on $L{\CalO}_{\xi}$ are the constant loops, taking values in the intersection points of the orbit ${\CalO}_{\xi}$ and a vector space
\begin{align*}
    ({\fg} / \bigoplus\limits_{\alpha} {\fg}_{\alpha})^{*} \cong {\mathfrak{t}}^{*}.
\end{align*}
Assuming (without loss of generality) that $\xi \in {\mathfrak{t}}^{*}$, we see that these intersection points are parametrized by elements of the Weyl group $W$:
\begin{align*}
    {\CalO}_{\xi} \cap {\mathfrak{t}}^{*} = \{w\xi: w \in W\}.
\end{align*}

Evaluating
the fixed contributions, using the $\zeta$-function regularization of infinite products -- the denominator of DH formula involves the product
of weights of ${\BT}$-action on $T_{w \xi}\left( L{\CalO}_{\xi} \right)$ -- one arrives at Weyl character formula \cite{HW}. 

In summary, the Wilson loop observable $W_{R}(C)$ in any gauge theory
arises from coupling that theory to an auxiliary quantum--mechanical system assigned to the loop $C$. 
\subsection{Wilson loop in Chern--Simons theory from disorder operator}

Chern--Simons theory is a three dimensional gauge theory with the gauge group $G$, the action $k S_{CS}$
for an integer $k$ and
\beq
S_{CS}(A) = \frac{1}{4\pi} \int_{M^3} {\Tr} \left( A dA + \frac{2}{3} A^{3} \right).
\label{eq:cs3}
\eeq
Space ${\CalH}_{\Sigma}$ of states on 
a spatial surface $\Sigma$, ${\rm dim}{\Sigma}=2$, is the space $H^{0}({\mathfrak{M}}_{\Sigma}, {\CalL}^{k})$ of holomorphic sections of the $k$-th power of the line bundle
${\CalL}$ over the moduli space ${\mathfrak{M}}_{\Sigma}$ of flat $G$-connections over ${\Sigma}$.

Now we insert
$W_{R}(C)$ in Chern--Simons path integral:
\begin{multline}
\langle W_{R}(C) \rangle = \int_{\rm gauge\, equivalence\, classes\, of\, connections}\, \left[ {\CalD}A \right]
\, e^{\ii k CS(A)} \, W_{R}(C) = \\
\int_{ \{{\rm gauge\ fields \ } A\} \times \{{\rm loops\ } \eta:\, C \rightarrow {\CalO}_{\xi} \}} \left[ {\CalD}A \right] \left[ {\CalD} {\eta} \right]\ e^{\frac{\ii k}{4\pi} \int_{M^{3}} {\Tr} \left( A dA + \frac{2}{3} A^{3} \right) + 2\pi\ii \oint_{C} d^{-1}{\omega}_{\xi}  + \ii \oint_{C} \left( {\eta}, A \right)}
\end{multline}
Now, the component of the gauge field $A$ along the loop acts as a Lagrange multiplier, imposing the local constraint:
\beq
F_{A} = \frac{k}{2\pi} {\eta} {\delta}^{(2)}_{C},
\label{eq:codim2sing}
\eeq
where $\delta^{(2)}_{C}$ is a two-form with delta-function support along the loop $C$. Thus we come to the notion of codimension $2$ vortex-type operators.
\subsection{Codimension 2 vortex-type operators in four dimensional Yang--Mills theory}
The three-dimensional Chern--Simons example above motivates the introduction of operators such as \eqref{eq:codim2sing}, but 
the most interesting setting for their study is
the four dimensional (super)-Yang--Mills theory. 
One can define the operators/observables associated with two dimensional surfaces $\Sigma$ embedded into
the four dimensional spacetime, as an instruction to perform the path integral over the space of gauge fields with a singularity along $\Sigma$, modelled
on \eqref{eq:codim2sing}. Now, $\eta$ will vary along $\Sigma$, while remaining inside $\CalO_{\xi}$. In other words, we have coupled the four dimensional gauge theory to a two dimensional sigma model valued in ${\CalO}_{\xi}$. Unlike the case of loop operators
in three dimensions, the two dimensional sigma model
valued in ${\CalO}_{\xi}$ comes with a plethora
of additional topological charges
\beq
{\vec k} \in {\pi}_{2}({\CalO}_{\xi})
\label{eq:topch}
\eeq
measuring the degree of the map of $\Sigma$ to ${\CalO}_{\xi}$. Thus, the surface defect in four dimensions depends on continuous parameters, the fugacities for \eqref{eq:topch}. 

Specifically for $SU(N)$ gauge theory, the orbits
${\CalO}_{\xi}$ are the flag varieties, so the sigma model in question describes the families of flags parametrized by $\Sigma$. As often happens in supersymmetric theories, the path integral localizes onto the solutions of some first order PDE, e.g.
on holomorphic maps of $\Sigma$ to the (generalized) flag variety. 

The moduli space of holomorphic maps is non-compact, its partial compactification, well-motivated by gauge theory, is the moduli space of parabolic sheaves. 

For the algebraic geometric approach relating parabolic bundles to orbifolds we use below, see \cite{biswas}.

This is the space we integrate over in the sections {\bf 7.3-7.4} below.

\subsection{$Y$-observable for $\hat{A}_{0}$-theory}
We use a common
\begin{nt}\normalfont
    \begin{align*}
         c_{ab}(\square = (i, j)) = \varepsilon_a (i-1) + \varepsilon_b (j-1), \; a, b = 1, ..., 4.
    \end{align*}
\end{nt}
\begin{defn}\normalfont
    Consider the $\hat{A}_{0}$ statistical model. The $Y$-observable is a function
    \begin{align*}
Y: \Pi^{N} \rightarrow &\{\text{rational functions of } x\} (H \text{ case})\\
&\{\text{meromorphic functions of } 1 - e^{-x}\} (K \text{ case})\\
&\{\text{meromorphic sections of $deg = N$ line bundle on $\mathcal{C}_{\sigma}$}\} (Ell \text{ case})
\end{align*}
\begin{align}\label{Y}
    Y(x)[\lambda^{(1)}, ..., \lambda^{(N)}] = \prod\limits_{\alpha = 1}^{N} \frac{\prod\limits_{\square \in \partial_{+}\lambda^{(\alpha)}}\theta(x - a_{\alpha} - c_{12}(\square))}{\prod\limits_{\square \in \partial_{-}\lambda^{(\alpha)}}\theta(x - a_{\alpha} - c_{12}(\square) - \varepsilon_1 - \varepsilon_2 )}
\end{align}
\end{defn}
\begin{rem}\normalfont
    The expectation value of \eqref{Y}
has poles at
\begin{align}\label{pole}
    x = a_{\alpha} + \varepsilon_1 (i - 1) + \varepsilon_2 (j - 1), \qquad (i, j) \in \bigcup\limits_{\alpha}\partial_{-}\lambda^{(\alpha)} \subset \mathbb{N}^2.
\end{align}
A pole corresponding to a particular box $\square = (i, j)$ comes from $\lambda^{(\alpha)}$'s whose inner boundary $\partial_{-} \lambda^{(\alpha)}$ contains $(i, j)$.
\end{rem}
\subsection{$\widehat{A_{0}}$-qq-character}
Let us denote by $\tilde{\lambda}^{(\alpha)}$ the diagram obtained from $\lambda^{(\alpha)}$ by removing $~\square = (i, j) \in \partial_{-}\lambda^{(\alpha)}$ (then $\square$ is also in $\partial_{+}\tilde{\lambda}^{(\alpha)}$). An important observation is that the function $Y(x)[\lambda^{(1)}, \ldots , \tilde{\lambda}^{(\alpha)}, \ldots , \lambda^{(N)}]$ vanishes at the point $(\ref{pole})$. Thus we can cancel this pole by subtracting a function proportional to $Y(x)^{-1}$. 
\begin{exer}
    Show that the expectation value of 
\begin{align*}
     Y(x) + \mathfrak{q} \frac{(\varepsilon_1 + \varepsilon_3)(\varepsilon_1 + \varepsilon_4)}{\varepsilon_3 \varepsilon_4} \frac{Y(x - \varepsilon_3)Y(x - \varepsilon_4)}{Y(x)} 
\end{align*}
has no poles at the points $(\ref{pole})$.
\end{exer}
Although the poles coming from expectation value of $Y(x)$ disappear, the added term brings new ones. Therefore the higher order terms of such a kind should be summed up leading to an infinite series. The result happens to be convergent.
\begin{pr}
The expectation value of observable
    \begin{align}\label{qq}
    \mathfrak{X}(x) = \sum\limits_{\lambda} \mathfrak{q}^{|\lambda|} &\prod\limits_{\square \in \lambda} 
    \frac{\theta(\varepsilon_3 (a_{\square} + 1) - \varepsilon_4 l_{\square} + \varepsilon_1 )\theta(-\varepsilon_3 a_{\square} + \varepsilon_4 (l_{\square} - 1) + \varepsilon_2)}{\theta(\varepsilon_3 (a_{\square} + 1) - \varepsilon_4 l_{\square})\theta(-\varepsilon_3 a_{\square} + \varepsilon_4 (l_\square + 1))} \nonumber \times\\
    &\qquad\qquad\qquad \times\ \frac{\prod\limits_{\square \in \partial_{+} \lambda}Y(x + \varepsilon_1 + \varepsilon_2 + c_{34}(\square))}{\prod\limits_{\square \in \partial_{-}\lambda}Y(x + c_{34}(\square))}
\end{align}
is holomorphic in the $x$-variable. This observable is called a fundamental $qq$-character for $\hat{A}_{0}$-theory (or order operator) (\cite{Nekrasov2015BPSCFTCN}).
\end{pr}
Absence of poles of the expectation value
\begin{align*}
    \langle \mathfrak{X}(x) \rangle = \frac{1}{Z^{inst}}\sum\limits_{(\lambda^{(\alpha)})} \mu_{(\lambda^{\alpha})}\, \prod\limits_{\alpha  = 1}^{N} {\qe}^{|{\lambda}^{({\alpha})}|}\,  \mathfrak{X}(x)[(\lambda^{(\alpha)})]
\end{align*}
allows one to represent $\langle \mathfrak{X}(x) \rangle = {\mathfrak{X}}_{0} \cdot \prod\limits_{\alpha = 1}^{N} \theta(x - A_{\alpha})$, where ${\mathfrak{X}}_{0}$ and
$A_{\alpha}$'s are some $\mathfrak{q}$-dependent constants
\begin{align*}
    A_{\alpha} = a_{\alpha} + \mathfrak{q}(...) \, , \ {\mathfrak{X}}_{0} = 1 + {\qe} (...)
\end{align*}
\subsection{The order operator as pushforward}
The formula \eqref{qq} has the interpretation in terms of a slightly more general statistical model, a particular instance of gauge origami \cite{Nekrasov:2016ydq}:

The measured space is the set of pairs of multi--partitions $\Pi^{N} \times \Pi^{M}$ equipped with the product measure $\mu_{(\lambda^{(\alpha)}) \times (\lambda^{(\beta)})} = \mu^{1}_{(\lambda^{(\alpha)})} \times \mu^{2}_{(\lambda^{(\beta)})}$. Each factor is the measure introduced earlier:
\begin{align*}
    \mu^{1} =  \mu(\vec{a}, \tau, {\ve}_{3};\varepsilon_1 , \varepsilon_2 ),\;\;\;
    \mu^{2} =  \mu(\vec{b}, \tau, {\ve}_{1};\varepsilon_3 , \varepsilon_4 ).
\end{align*}

\begin{pr}
(\cite{Nekrasov2015BPSCFTCN}) Define the observable
\begin{align*}
\tilde{\mathfrak{X}}: \Pi^{N} \times \Pi^{M} \rightarrow \begin{cases} \{\text{rational functions of } x\} \, , \qquad & H \ \text{ case}\\
\{\text{meromorphic functions of } \ e^{x}\} \, , \qquad & K \ \text{ case}\\
\{\text{meromorphic sections of $deg = N$ line bundle on $\mathcal{C}_{\sigma}$}\} \, ,  \qquad & Ell \text{ case} \end{cases}
\end{align*}
\begin{align*}
  &\mathfrak{\tilde{X}}(x)[(\lambda^{(\alpha)}), (\lambda^{\beta})] =  \prod\limits_{\alpha = 1}^{N} \prod\limits_{\beta = 1}^{M}
  \prod\limits_{\stackrel{\square \in \partial_{+}\lambda^{(\alpha)}}{\blacksquare \in \partial_{-} \lambda^{(\alpha)}}}
   \prod\limits_{\stackrel{\square' \in \partial_{+}\lambda^{(\beta)}}{\blacksquare' \in \partial_{-} \lambda^{(\beta)}}}\\
   &\frac{\theta\big(x - a_{\alpha} + b_{\beta} + \varepsilon_{1} + \varepsilon_{2} + c_{34}(\square) - c_{12}(\square')\big)\theta\big(x - a_{\alpha} + b_{\beta} - \varepsilon_1 - \varepsilon_2 + c_{34}(\blacksquare) - c_{12}(\blacksquare')\big)}{\theta\big(x - a_{\alpha} + b_{\beta} + c_{34}(\blacksquare) - c_{12}(\square')\big)\theta\big(x - a_{\alpha} + b_{\beta} + c_{34}(\square) - c_{12}(\blacksquare')\big)}.
\end{align*}
Then the observable $\mathfrak{X}$ $(\ref{qq})$ is the result of integrating $\tilde{\mathfrak{X}}$ along the fibers of projection onto the first factor:
\begin{align*}
  \mathfrak{X} = \pi_{1*}\tilde{\mathfrak{X}}, \;\;\; \pi_1 : \Pi^{N} \times \Pi^{M} \rightarrow \Pi^{N}.
\end{align*}
\end{pr}
\subsection{Disorder operators: Surface defects}
Analogously to Wilson loop operators in Chern--Simons theory one can explore codimension-two defects in 4d Yang--Mills (\cite{Nekrasov:2021tik}). This leads to the study of sheaves equipped with parabolic structure. For example, for a sheaf $\mathcal{E}$ on $\mathbb{C}^{2}$ a parabolic structure along the $D = \{z_{2} = 0\} \subset \mathbb{C}^{2}$-plane is the filtration with subsheaves
\begin{align}
    \mathcal{E} = \mathcal{E}_0 \supset \mathcal{E}_{1} \supset ... \supset \mathcal{E}_{N-1} \supset \mathcal{E}_{N} = \mathcal{E}(-D).
\end{align}
Integration over the moduli spaces of sheaves with parabolic structure produces a new statistical model with a different measure compared with $(\ref{A0hatm})$. In this section we give its description. 

Choose an arbitrary generator $\Omega$ of the cyclic group $\Gamma = \mathbb{Z}/N\mathbb{Z}$. We identify the set of irreducible representations $\hat{\Gamma}$ of $\Gamma$ with the set $\{0, ..., N-1\}$ by $\omega \mapsto \frak{R}_{\omega}$,
where
\begin{align*}
    \mathfrak{R}_{\omega}: \Gamma &\rightarrow \mathbb{C}^{*}\\
    \Omega &\mapsto e^{\frac{2 \pi {\ii}\omega}{N}}.
\end{align*}
It allows us to make no distinction between $\hat{\Gamma}$ and $\{0, .., N-1\}$ in the following. Let us fix an arbitrary bijection $c: \{1, ..., N\} \rightarrow \{0, ..., N-1\}$.
For each multi--partition $(\lambda^{(\alpha)})$ we dye its every Young diagram $\lambda^{(\alpha)}$ in colors of $\hat{\Gamma}$.
Boxes in the $j$'s column have by definition the same color $c(\alpha) + j - 1 \mod N \in \mathbb{Z}/N\mathbb{Z}$.
For every $\omega \in \{0, ..., N-1\}$ we define an observable $k_{\omega} = k_{\omega}[(\lambda^{(\alpha)})]$ as the total length
\begin{align*}
    k_{\omega} = \sum\limits_{\alpha = 1}^{N} \#\{(i, j) \in \lambda^{(\alpha)}:c(\alpha) + j - 1 \equiv \omega \mod N\}
\end{align*}
of the $\omega$-colored columns. Instead of a single fugacity $\mathfrak{q}$ we now have $N$ of them
    $\{ \mathfrak{q}_{\omega}, ~ \omega = 0, ..., N-1 \}$.
    Let $L \subset {\BC}$ denote the lattice
    \beq
    L = \mathbb{Z}\varepsilon_{1} +  \mathbb{Z}\varepsilon_{2} + 
   \mathbb{Z}\varepsilon_{3} +  \mathbb{Z}\varepsilon_{4} +  \sum\limits_{\alpha = 1}^{N} \mathbb{Z}a_{\alpha}
   \eeq
We use the $\mathbb{Z}$-linear map $ [\;\;]: L \rightarrow \mathbb{Z}/N\mathbb{Z}$ defined by
\begin{align*}
    [a_{\alpha}] &= c(\alpha)\\
   [\varepsilon_{1}] &= 0\\
      [\varepsilon_{2}] &= 1\\
         [\varepsilon_{3}] &= 0\\
            [\varepsilon_{4}] &= -1
\end{align*}
and a projection operator $f \mapsto f^{\delta}$, acting on functions $f:L \rightarrow \mathbb{C}$:
\begin{align}\label{deltadef}
    f^{\delta}(x) = 
    \begin{cases}
        f(x), \ & [x] = 0 \\
       \  1, \ &  \text{otherwise}.
    \end{cases}
\end{align}
\begin{defn}\normalfont
    Define a measure
    \begin{align*}
    \mu^{\rm orb}_{(\lambda^{(\alpha)})} = \prod\limits_{\omega = 0}^{N-1} \mathfrak{q}_{\omega}^{k_{\omega}}\prod\limits_{\alpha , \beta = 1}^{N} &\prod\limits_{(i, j) \in \lambda^{(\beta)}}  \frac{\theta^{\delta} \left( \varepsilon_3 + \varepsilon_1 (i - \lambda_{j}^{(\beta)t}) + \varepsilon_2 \left( 1 + \lambda_{i}^{(\alpha)} - j \right) + a_{\alpha} - a_{\beta} \right)}{\theta^{\delta} \left( \varepsilon_1 \left(i - \lambda_{j}^{(\beta)t}\right) + \varepsilon_2 \left(1 + \lambda_{i}^{(\alpha)} - j\right) + a_{\alpha} - a_{\beta} \right)} \ \times\\
     &\prod\limits_{(i, j) \in \lambda^{(\alpha)}} \frac{\theta^{\delta} \left( \varepsilon_3 + \varepsilon_1 (\lambda_{j}^{(\alpha)t} + 1 - i) + \varepsilon_2 \left( j -\lambda_{i}^{(\beta)} \right) + a_{\alpha} - a_{\beta}\right)}{\theta^{\delta}\left(\varepsilon_1 \left(\lambda_{j}^{(\alpha)t} + 1 - i \right)  + \varepsilon_2 \left( j -\lambda_{i}^{(\beta)} \right) + a_{\alpha} - a_{\beta}\right)}
\end{align*}
and a map
\begin{align*}
    & \pi \, :\, \Pi^{N} \ \longrightarrow \Pi^{N}\, , \\
  {\pi}\, : \,  (\lambda^{(\alpha)}) &\mapsto (\Lambda^{(\alpha)})\, , \;\; \Lambda_{j}^{(\alpha)t} = \lambda_{\alpha + N(j - 1)}^{(\alpha)t}.
\end{align*}
A $\textit{surface defect (disorder) operator}$ $\mathcal{S}$ is a density of ${\pi}_{*}\mu^{\rm orb}$ with respect to the measure $\mu$ \eqref{A0hatm}:
\begin{align*}
    \pi_{*}\mu^{\rm orb} &= \mu( {\bf a}, \tau , {\ve}_{3} ;\ve_1 , \ve_2 ) \times \mathcal{S}(\vec{w}),\\ 
    e^{2 \pi {\ii} \tau} &= \mathfrak{q} = \prod\limits_{\omega \in \hat{\Gamma}}\mathfrak{q}_{\omega}\, , \ 
    \vec{w} = [w_0: ... : w_{N}]\\
    \frac{w_1}{w_0} &= \mathfrak{q}_{0}\, , \, \ldots  \, , \,  \frac{w_{N-1}}{w_{N-2}} = \mathfrak{q}_{N-2}\, , \, \frac{w_{N}}{w_{N-1}} = \mathfrak{q}^{-1}\mathfrak{q}_{N-1}.
\end{align*}
\end{defn}
\begin{rem}\normalfont
    The observable $\mathcal{S}(\vec{w}) = \mathcal{S}(\vec{w}) [(\Lambda^{\alpha})]$ can be constructed by integration
\begin{align*}
    \mathcal{S}(\vec{w}) [(\Lambda^{\alpha})] = \sum\limits_{k_{\omega} \geq 0} \prod\limits_{\omega \in \hat{\Gamma}}\mathfrak{q}_{\omega}^{k_{\omega}} \int\limits_{\mathcal{M}_{(k_{\omega})}(\Lambda)} 1
\end{align*}
over the moduli spaces $\mathcal{M}_{(k_{\omega})}(\Lambda)$ of (maximal) parabolic structures $(\mathcal{E}_{\omega})$ for the fixed sheaf $\mathcal{E}$ specified by $\Lambda$ and fixed first Chern classes $c_{1}$ of the consecutive factors $(\mathcal{E}_{\omega}/\mathcal{E}_{\omega + 1})$ specified by $k_{\omega}$.
\end{rem}

\section{Indirect gauge theory --- integrable systems correspondence}

In this section we illustrate, through a particular example, the relation between the statistical models arising from gauge theories and integrable systems. Among the many applications of this relation are the classical/quantum correspondence, including the Bethe ansatz, and the connection to the geometric Langlands program, first observed in \cite{Feigin:1994in}.
 We refer the reader to \cite{Etingof:2023drx, Jeong:2023qdr, Jeong:2024hwf} for modern developments and their place in the BPS/CFT correspondence.

We consider a particular version of $\hat{A}_{r}$-theory, 
which generalizes the previously discussed $\hat{A}_{0}$-theory.

\subsection{$\hat{A}_{r+1}$-theory with surface defect}
We define a statistical mechanical model and introduce corresponding fundamental $qq$-characters.

Measured space:
\begin{align*}
    \Pi^{N_0}\times \ldots \times \Pi^{N_{r+1}} = \left\{\, \left( \lambda^{(\alpha, s)} \right)_{\alpha = 1, ..., N_{s}}^{s = 0, ..., r+1}\, \Biggr| \ {\lambda}^{(\alpha, s)} \in \Pi \, \right\}.
\end{align*}
The measure $\mu = \mu_{(\lambda^{(\alpha, s)})}$ depends on additional parameters $((a_{\alpha, s}), \vec{\mathfrak{q}}, \varepsilon_1, \varepsilon_2, \varepsilon_3)$, where
\begin{align*}
    &(a_{\alpha, s})_{\alpha =1, ..., N_{s}}^{s = 0, ..., r+1} \in \mathbb{C}^{N_{0}} \times ... \times \mathbb{C}^{N_{r+1}},\\
    &\vec{\frak{q}} = (\frak{q}_{0}, ..., \frak{q}_{r+1}) - r+2 \text{ fugacities},\\
    &(\varepsilon_1, ..., \varepsilon_{4}) \in \mathbb{C}^{4},
\end{align*}
\begin{align*}
    \tilde{\varepsilon}_1 = \varepsilon_1 , \;\;\;\; \tilde{\varepsilon}_2 = \frac{\varepsilon_2}{p}, \;\;\;\; \tilde{\varepsilon}_3
 = \frac{\varepsilon_3}{r + 2}, \;\;\;\; \tilde{\varepsilon}_4 = - \tilde{\varepsilon}_1 - \tilde{\varepsilon}_2 - \tilde{\varepsilon}_3 ,
 \end{align*}
 \begin{align*}
    c: \{(\alpha, s): s = 0, ..., r+1, \; \alpha = 1, ..., N_s\} \rightarrow \mathbb{Z}/p\mathbb{Z}.
\end{align*}
Assigning:
\begin{align*}
    [\;\;] = ([\;\;]', [\;\;]''):\mathbb{Z}\varepsilon_{1} \oplus \mathbb{Z}\tilde{\varepsilon}_{2}\oplus 
   &\mathbb{Z}\tilde{\varepsilon}_{3} \oplus \mathbb{Z}\tilde{\varepsilon}_{4}\oplus \bigoplus\limits_{\stackrel{\alpha = 1, ..., N_{s}}{s = 0, ..., r+1}} \mathbb{Z}a_{\alpha, s} \rightarrow \mathbb{Z}/(r+2)\mathbb{Z}\times \mathbb{Z}/p\mathbb{Z}\\
    [a_{\alpha, s}] &= (s, c(\alpha, s))\\
   [\varepsilon_{1}] &= (0, 0)\\
      [\tilde{\varepsilon}_{2}] &= (0, 1)\\
         [\tilde{\varepsilon}_{3}] &= (1, 0)\\
            [\tilde{\varepsilon}_{4}] &= (-1, -1)
\end{align*}
For every $s \in {0, ..., r+1}$ we define an observable 
\begin{align*}
    k_s = \sum\limits_{\alpha = 1}^{N_{s}} |\lambda^{(\alpha, s)}|.
\end{align*}

The measure is given by the formula
\begin{align*}
    \mu_{(\lambda^{(\alpha, s)})} = \prod\limits_{s = 0}^{r+1} \frak{q}_{s}^{k_s} \prod\limits_{s, l = 0}^{k+1} \prod\limits_{\alpha = 1}^{N_{s}}\prod\limits_{\beta = 1}^{N_{l}} &\prod\limits_{(i, j) \in \lambda^{(\beta, l)}} \frac{\theta^{\delta} (\varepsilon_3 + \varepsilon_1 (i - \lambda_{j}^{(\beta, l)t}) + \varepsilon_2 (1 + \lambda_{i}^{(\alpha, s)} - j) + a_{\alpha, s} - a_{\beta, l})}{\theta^{\delta}(\varepsilon_1 (i - \lambda_{j}^{(\beta, l)t}) + \varepsilon_2 (1 + \lambda_{i}^{(\alpha, s)} - j) + a_{\alpha, s} - a_{\beta, l})} \\
    &\prod\limits_{(i, j) \in \lambda^{(\alpha, s)}} \frac{\theta^{\delta}(\varepsilon_3 + \varepsilon_1 (\lambda_{j}^{(\alpha, s)t} + 1 - i) + \varepsilon_2 (j -\lambda_{i}^{(\beta, l)}) + a_{\alpha, s} - a_{\beta, l})}{\theta^{\delta}(\varepsilon_1 (\lambda_{j}^{(\alpha, s)t} + 1 - i) + \varepsilon_2 (j -\lambda_{i}^{(\beta, l)}) + a_{\alpha, s} - a_{\beta, l})}
\end{align*}
Introduce the following functions
\begin{align*}
    m_{\lambda}(\varepsilon_1 , ..., \varepsilon_{4}) = \prod\limits_{\square = (i, j) \in \lambda} S^{\delta}(-\tilde{\varepsilon}_{3} l_{\square} + \tilde{\varepsilon}_{4} (a_{\square} + 1)),
\end{align*}
\begin{align*}
    S(x) = \frac{\theta(x + \varepsilon_1) \theta(x + \tilde{\varepsilon}_2)}{\theta(x) \theta(x + \tilde{\varepsilon}_1 + \tilde{\varepsilon}_2 )},
\end{align*}
where $S^{\delta}$ is defined by $(\ref{deltadef})$.
There are $(r+2) \times p$ fundamental $qq$-characters:
 \begin{align}\label{genqq}
     \mathfrak{X}_{\omega, s}(\varepsilon_1 , ... , \varepsilon_4 )[(\lambda^{(\alpha , s)})](x) = \sum\limits_{\lambda} m_{\lambda}(\varepsilon_1 , ... , \varepsilon_4 ) \mathfrak{X}_{\lambda}(\omega, s, x, (\lambda^{(\alpha, s)}), \varepsilon_1 , ..., \varepsilon_4 ),
 \end{align}
 where $\mathfrak{X}_{\lambda}(\omega, s, x, (\lambda^{(\alpha, s)}), \varepsilon_1 , ..., \varepsilon_4 )$ are some explicit (but slightly complicated) expressions of parameters. In the following we focus on the $A_{r}$-theory, which may be viewed as a degeneration of $\hat{A}_{r}$-theory.
\subsection{$A_{r}$ - theory}
This is a special case of the previous system for
\begin{align*}
    \mathfrak{q}_{0} &= \mathfrak{q}_{r+1} = 0,\\
    N_{0} &= ... = N_{r+1} = N.
\end{align*}
Let us first consider the case without surface defect, which corresponds to $p = 1$. The following notation is convenient:
\begin{align*}
    &m_{\alpha}^{-} = a_{\alpha, 0},\\
    &m_{\alpha}^{+} = a_{\alpha, r+1} + \varepsilon_{1} + \varepsilon_{2},\\
    &\vec{z} = (z_0, ..., z_{r}), \\
    &z_{i} = z_0 \mathfrak{q}_{1} ... \mathfrak{q}_{i}, \; i = 1, ..., r.
\end{align*}
Let us define the so-called $Y$-observables:
\begin{align*}
 &Y_{s}(x)[(\lambda^{(\beta, l)})] = \prod\limits_{\alpha = 1}^{N} \frac{\prod\limits_{\square \in \partial_{+}\lambda^{(\alpha, s)}} \theta (x - a_{\alpha, s} - c_{12}(\square))}{\prod\limits_{\square \in \partial_{-}\lambda^{(\alpha, s)}} \theta (x - a_{\alpha, s} - \varepsilon_1 - \varepsilon_2 - c_{12}(\square))},\\
      &Y_{0}(x) = \prod\limits_{\alpha = 1}^{N} \theta(x - m_{\alpha}^{-}),\\
    &Y_{r+1}(x-\varepsilon_1 - \varepsilon_2 ) = \prod\limits_{\alpha = 1}^{N}\theta (x - m_{\alpha}^{+}).
\end{align*}
It is also useful to introduce the observables
\begin{align*}
    \bold{\Lambda}_{i}(x) = z_{i} \frac{Y_{i+1}(x + \varepsilon_1 + \varepsilon_2)}{Y_{i}(x)}, \; i = 0, ..., r.
\end{align*}
The general formula $(\ref{genqq})$ for $l$'th $qq$-character $(l \in \{1, ..., r+1\})$ in this case simplifies (\cite{Nekrasov2015BPSCFTCN}) to
\begin{align}\label{Aqq}
    \mathfrak{X}_{l}(x) = Y_{0}(x + (\varepsilon_1 + \varepsilon_2 )(1-l))\sum\limits_{0 \leq i_1 < ... <i_{l} \leq r} \prod\limits_{b = 1}^{l} \frac{\bold{\Lambda}_{i_{b}}(x + (\varepsilon_1 + \varepsilon_2 )(b-1))}{z_{b - 1}}.
\end{align}
\begin{exer}
    Compute the first two terms of $(\ref{Aqq})$:
    \begin{align*}
        \mathfrak{X}_{l} = Y_{l}(x + \varepsilon_1 + \varepsilon_2 ) + \mathfrak{q}_{l}\frac{Y_{l-1}(x)Y_{l+1}(x + \varepsilon_1 + \varepsilon_2)}{Y_{l}(x)}+...
    \end{align*}
\end{exer}
The very nice property of (normalized) expectation value 
\begin{align}\label{expvalqq}
    \langle \mathfrak{X}_{l}(x) \rangle = \frac{1}{Z^{inst}}\sum\limits_{(\lambda^{(\alpha, s)})} \mu_{(\lambda^{\alpha, s})} \mathfrak{X}_{l}(x)[(\lambda^{(\alpha, s)})]
\end{align}
of our observable is that it has no poles with respect to the variable $x$, hence in elliptic case $\langle \mathfrak{X}_{l}(x) \rangle$ is a holomorphic section of degree $N$ line bundle on elliptic curve and in rational case it is a degree $N$ polynomial.
\subsection{Spectral curve}
Classical integrable system arises after taking $(\textit{thermodynamic})$ limit $\varepsilon_1 , \varepsilon_2 \rightarrow 0$. 
Let us discuss this limit for expectation value $(\ref{expvalqq})$:
\begin{align*}
    Y_{l}'(x) = \lim\limits_{\varepsilon_1 , \varepsilon_2 \rightarrow 0} \langle Y_{l}(x) \rangle,\\
    T_{l}(x) = \lim\limits_{\varepsilon_1 , \varepsilon_2 \rightarrow 0}  \langle \mathfrak{X}_{l}(x) \rangle.
\end{align*}

In fact we have the factorization property asymptotically
\begin{align*}
    \langle Y_{l}(x)^{-1} \rangle \sim \frac{1}{ Y_{l}'(x)}, \;\;\; \Lambda_{i}(x) = \langle \bold{\Lambda}_{i}(x) \rangle \sim z_i  \frac{ Y_{i+1}'(x) }{  Y_{i}'(x) } , \;\;\; \langle \prod\limits_{b = 1}^{l}\bold{\Lambda}_{i_{b}}(x) \rangle \sim \prod\limits_{b = 1}^{l}\Lambda_{i_b}(x),
\end{align*}
and thus
\begin{align*}
    T_{l}(x) = \frac{Y_{0}'(x) }{z_0 ... z_{l-1}}e_{l}(\Lambda_{0}(x), ..., \Lambda_{r}(x)),
\end{align*}
where $e_{l}$'s are elementary symmetric functions. Using the defining identity for symmetric functions
\begin{align*}
    \prod\limits_{i = 0}^{r}(1 - z^{-1}\Lambda_{i}) = \sum\limits_{l = 0}^{r + 1}(-z)^{-l}e_{l}(\Lambda_{0}, ..., \Lambda_{r}),
\end{align*}
we conclude
\begin{align*}
   R_{\vec{z}}(x, z) \overset{def}{=} Y_{0}'(x) + \sum\limits_{s = 1}^{r + 1}z_{0}...z_{s-1}T_{s}(x)(-1/z)^{s} = Y_{0}'(x)\prod\limits_{i = 0}^{r}\left( 1 - z^{-1} z_{i}\frac{ Y_{i+1}'(x) }{ Y_{i}'(x)} \right).
\end{align*}
This function of two variables defines the spectral curve
\begin{align*}
    \mathcal{C}_{\vec{z}} = \{R_{\vec{z}}(x, z) = 0\} \subset E_{\mathfrak{q}} \times \mathbb{CP}^{1},
\end{align*}
where $z$ is a coordinate on $\mathbb{CP}^{1}$. To see the Lax operator we need to introduce a surface defect.
\subsection{Lax operator}
With presence of surface defect one also has $Y$-observables $Y_{s,\omega}(x)$ and $qq$-characters $\mathfrak{X}_{s, \omega}(x)$. It is convenient to organize them in $r+2$ diagonal matrices of size $N \times N$
\begin{align*}
    \hat{Y}_{s}(x) = \mathrm{diag} \big( Y_{s, 0}(x), ..., Y_{s, N-1}(x) \big),\\
    \hat{\mathfrak{X}}_{s}(x) = \mathrm{diag} \big( \mathfrak{X}_{s, 0}(x), ..., \mathfrak{X}_{s, N-1}(x) \big)
\end{align*}
and introduce the notation
\begin{align*}
\hat{Z}_{i} &= \mathrm{diag} \big( z_{i,0} , ..., z_{i, N-1} \big),\\
     z_{i, \omega} &= z_{0,\omega} \mathfrak{q}_{1, \omega} ... \mathfrak{q}_{i, \omega}, \;\; i = 0, ..., r.
\end{align*}
Now taking the limit
\begin{align*}
    &\hat{Y}_{l}'(x) = \lim\limits_{\varepsilon_1 , \varepsilon_2 \rightarrow 0} \langle \hat{Y}_{l}(x) \rangle,\\
    &\hat{T}_{l}(x) = \lim\limits_{\varepsilon_1 , \varepsilon_2 \rightarrow 0}  \langle \hat{\mathfrak{X}}_{l}(x) \rangle ,\\
    &\hat{\Lambda}_{i}(x) = \lim\limits_{\varepsilon_1 , \varepsilon_2 \rightarrow 0} \hat{Z}_{i} \langle \frac{\hat{Y}_{i+1}(x)}{\hat{Y}_{i}(x)}\rangle
\end{align*}
we define a matrix-valued function
\begin{align*}
    \hat{D}(z, x) = \hat{Y}_{0}'(x)\prod\limits_{i = 0}^{\overset{\longrightarrow}{r}}\big( 1 - \hat{C}_{z} \hat{\Lambda}_{i}(x)\big) = : \hat{Y}_{0}'(x) + \sum\limits_{s = 1}^{r+1}\prod\limits_{i = 0}^{\overset{\longrightarrow}{s-1}}(-\hat{C}_{z}\hat{Z}_{i}) \hat{T}_{s}(x),
\end{align*}
where
\begin{align*}
    \hat{C}_{z} = 
    \begin{pmatrix}
        0 & 1\\
          & 0 & 1\\
          &   & \cdot & \cdot\\
          &   &  & \cdot & \cdot\\
          &   &  &  & 0 & 1\\
         z^{-1} &  &  &  &      & 0  
    \end{pmatrix}.
\end{align*}

\begin{pr}\label{dets}
The following identities hold
\begin{align*}
z_{i} = \det \hat{Z}_{i}, \;\;\;
Y_{i}'(x) = \det \hat{Y}_{i}'(x), \;\;\;
\det(\hat{D}(z, x)) = R(z, x).
\end{align*}
\end{pr}
We perform the Lax operator for rational variant of the theory. Note that in this case $\hat{D}$ is a matrix valued polynomial of degree $1$ as a function of $z$ and hence might be written as
\begin{align*}
    \hat{D}(z, x) = \hat{D}_{0}x + \hat{D}_{1} 
\end{align*}
for some $x$-independent constants $\hat{D}_{0}$ and $\hat{D}_{1}$. $\hat{D}_{0}$ can be easily extracted from $\hat{D}(z, x)$ by taking the limit $x \rightarrow \infty$:
\begin{align*}
    \hat{D}_{0} = \prod\limits_{i = 0}^{\overset{\longrightarrow}{r}}\big( 1 - \hat{C}_{z} \hat{Z}_{i}\big).
\end{align*}
We define a Lax operator by the formula
\begin{align*}
    \hat{L}(z) = -\hat{D}_{0}^{-1}(z) \hat{D}_{1}(z).
\end{align*}
From proposition $(\ref{dets})$ we deduce that the spectral curve is given by the equation
\begin{align*}
    \det (x - \hat{L}(z)) = 0.
\end{align*}
Furthermore, using a simple
\begin{statement}{Problem 8}\normalfont \label{prob8}
 Let $A$ be a diagonal matrix. Show that
    \begin{align*}
        \det(1 - \hat{C}_{z}A) = 1 - z^{-1}\det A .
    \end{align*}
\end{statement}
We see that $\hat{L}(z)$ has poles at $z = z_{i},\; i = 0, ..., r+1$ and therefore has a form
\begin{align*}
    \hat{L} = \sum\limits_{i = 0}^{r+1}\frac{\hat{L}_{i}}{z - z_i }.
\end{align*}
\begin{statement}{Problem 9}\normalfont \label{prob9}
 Prove that the residues $\hat{L}_{i}$ are rank $1$ operators.
\end{statement}
This is the Lax operator of Garnier--Gaudin system. The matrix valued $1$-form $\phi(z) =  \hat{L}(z)\frac{d z}{z}$ is the Higgs field  for Hitchin system on $\mathbb{CP}^{1}$ with $r+3$ punctures with general residues at $0$ and $\infty$ and rank $1$ residues at $z_i$'s. Note that there is a similar construction for $\hat{A}_{0}$-theory involving infinite products (\cite{Grekov:2023fek}, \cite{Grekov:2024fyh}). The corresponding $\hat{L}(z)$ is the Lax operator of elliptic Calogero--Moser system. 
\section{Solution of problems}
\subsection{Problem \myhref{prob1}{1}}
Let us use notation $L = L^{(1)} + L^{(2)}$ and $A = A^{(1)} + A^{(2)}$ for
\begin{align*}
    &L^{(1)}_{ij} = p_i \delta_{ij}, \;\;\;L^{(2)}_{ij}=(1 - \delta_{ij}) \frac{{\ii}\nu}{x_i - x_j}, \\
    &A^{(1)}_{ij} = (1 - \delta_{ij})\frac{{\ii}\nu}{(x_i - x_j)^2}, \;\;\; A^{(2)}_{ij} = - \delta_{ij} \sum\limits_{k \neq i} \frac{{\ii}\nu }{(x_i - x_k)^2}.
\end{align*}
Here and further $(1 - \delta_{ij})(\text{something that is not defined when } i=j)$ is equal to $0$ when $i=j$ by definition.

We compute each commutator separately
\begin{align*}
    [A^{(1)}, L^{(1)}]_{ij} = (1 - \delta_{ij})\frac{{\ii}\nu (p_j - p_i)}{(x_i - x_j)^2},
\end{align*}
\begin{align*}
    [A^{(1)}, L^{(2)}]_{ij} = -\nu^2 \sum\limits_{k} \frac{(1- \delta_{ik})(1 - \delta_{kj})}{(x_i - x_k)(x_k - x_j)}\left(\frac{1}{x_i - x_k} - \frac{1}{x_k - x_j}\right).
\end{align*}
Considering cases
\begin{align*}
\frac{1}{(x_i - x_k)(x_k - x_j)} =
    \begin{cases}
        -\frac{1}{(x_i - x_k)^{2}},\;\;\; i = j\\
        \frac{1}{x_i - x_j}\left( \frac{1}{x_i - x_k} + \frac{1}{x_k - x_j} \right) ,\;\;\;i \neq j
    \end{cases}
\end{align*}
we continue
\begin{align*}
    [A^{(1)}, L^{(2)}]_{ij} = 2\nu^2 \delta_{ij} \sum\limits_{k \neq i} \frac{1}{(x_i - x_k)^3} - \frac{\nu^2 (1 - \delta_{ij})}{x_i - x_j} \sum\limits_{k \neq i, j}\left( \frac{1}{x_i - x_k} + \frac{1}{x_k - x_j} \right).
\end{align*}
\begin{align*}
    [A^{(2)}, L^{(1)}]_{ij} = 0.
\end{align*}
\begin{align*}
    [A^{(2)}, L^{(2)}] =  \frac{\nu^2 (1 -\delta_{ij})}{x_i - x_j} \sum\limits_{k \neq i, j}\left( \frac{1}{x_i - x_k} + \frac{1}{x_k - x_j} \right).
\end{align*}
Puttting it altogether we get
\begin{align}\label{com}
    [A, L]_{ij} = [A^{(1)}, L^{(1)}]_{ij} = (1 - \delta_{ij})\frac{{\ii}\nu (p_j - p_i)}{(x_i - x_j)^2} + 2\nu^2 \delta_{ij} \sum\limits_{k \neq i} \frac{1}{(x_i - x_k)^3}.
\end{align}
Now compute derivative
\begin{align}\label{der}
    \dot{L}_{ij} = \dot{p}_{i} \delta_{ij} + (1-\delta_{ij}){\ii}\nu \frac{\dot{x}_{j} - \dot{x}_{i}}{(x_i - x_{j})^2}.
\end{align}
Equating $(\ref{com})$ and $(\ref{der})$ gives equations of motion \eqref{eq:EOMcm}.
\subsection{Problem \myhref{prob2}{2}}
Computing
\begin{align*}
    [P, X]_{ij} = (x_j - x_i)P_{ij}
\end{align*}
we see that condition $\mu = 0$ has the form
\begin{align*}
    (x_j - x_i )P_{ij} + {\ii} z_i z_j - {\ii} \nu \delta_{ij} = 0.
\end{align*}
With help of condition $z_i \in \mathbb{R}_{+}$ this (for $i=j$) implies $z_i = \sqrt{\nu}$. $P_{ij}$ for $i \neq j$ is also fixed by $\displaystyle P_{ij} = \frac{{\ii}\nu}{x_i - x_j} \displaystyle$. The $(\ref{Pmatrix})$ now follows from the choice $P_{ii} = p_i$.

\subsection{Problem \myhref{prob3}{3}}
We need to find $E_{ij}(z)$ satisfying the conditions
\begin{align}
\begin{cases}\label{mmc}
    \partial_{\bar{z}} E_{ij} - (\alpha_i - \alpha_j ) E_{ij} + (\delta_{ij} \nu - v_i w_j ) 2 \pi {\ii} \delta^{(2)} (z, \bar{z}) = 0 \\
    E_{ij}(z+1) = E_{ij} (z + \tau) = E_{ij} (z)
    \end{cases}
\end{align}
In case $i = j$ it means that $E_{ii}(z)$ is meromorphic function on $C_{\tau}$ with only possible first order pole at $z = 0$. It can only be a constant, hence
\begin{align*}
\begin{cases}
    E_{ii}(z) = E_{ii}(0) = p_i\\
    \nu - v_i w_i = 0 \stackrel{w_i = v_i}{\Rightarrow} v_i = \sqrt{\nu}.
\end{cases}
\end{align*}
In case $i \neq j$ we can introduce
\begin{align}\label{elax}
    L_{ij}(z) = e^{(\alpha_i - \alpha_j)(z - \bar{z})} E_{ij},
\end{align}
which because of $(\ref{mmc})$ should be meromorphic function satisfying
\begin{align*}
    \begin{cases}
        L_{ij} (z+1) = L_{ij}(z)\\
        L_{ij}(z + \tau) = e^{2 \Im \tau (\alpha_i - \alpha_j) } L_{ij} (z)\\
        L_{ij}(z) \underset{z \rightarrow 0}{\sim} \frac{\nu}{z} 
    \end{cases}
\end{align*}
In fact $L_{ij}$ is uniquely defined by these constraints and so we come up to a Krichever's Lax matrix for the Elliptic Calogero--Moser system \cite{Kr}
\begin{align*}
    L_{ij}(z) = \nu \frac{\theta' (0)}{\theta (x_i - x_j )}\frac{\theta(z + x_i - x_j)}{\theta(z)}, \; i \neq j ; \;\;L_{ii} = p_i . 
\end{align*}
In this way $(x_i , p_i)$ is a holomorphic coordinate system on the chosen leaf in $\mu^{-1}_{\infty}(0)$.

\subsection{Problem \myhref{prob4}{4}}

A contribution to the $Z^{inst}_{1}$ is made by the collections of partitions of type
\begin{align*}
(\lambda^{(1)}, ..., \lambda^{(N)}) = (\varnothing, ..., \varnothing, \stackrel{\alpha}{\square}, \varnothing, ..., \varnothing).
\end{align*}
For such a collection the product
\begin{align*}
\prod\limits_{\beta , \beta' = 1}^{N} &\prod\limits_{(i, j) \in \lambda^{(\beta')}} \frac{\varepsilon_3 + \varepsilon_1 (i - \lambda_{j}^{(\beta')t}) + \varepsilon_2 (1 + \lambda_{i}^{(\beta)} - j) + a_{\beta} - a_{\beta'}}{\varepsilon_1 (i - \lambda_{j}^{(\beta')t}) + \varepsilon_2 (1 + \lambda_{i}^{(\beta)} - j) + a_{\beta} - a_{\beta'}} \\
    &\prod\limits_{(i, j) \in \lambda^{(\beta)}} \frac{\varepsilon_3 + \varepsilon_1 (\lambda_{j}^{(\beta)t} + 1 - i) + \varepsilon_2 (j -\lambda_{i}^{(\beta')}) + a_{\beta} - a_{\beta'}}{\varepsilon_1 (\lambda_{j}^{(\beta)t} + 1 - i) + \varepsilon_2 (j -\lambda_{i}^{(\beta')}) + a_{\beta} - a_{\beta'}}
\end{align*}
contains the factors of different types distinguished by the values of $\beta$ and $\beta'$. If neither $\beta$ nor $\beta'$ equals to $\alpha$ then the factor is unit. If only of one of the $\beta, \beta'$ equals to $\alpha$ then the factor is given by
\begin{align*}
&\frac{\varepsilon_{3} + \varepsilon_{1} + \varepsilon_{2} + a_{\alpha \beta'}}{\varepsilon_{1} + \varepsilon_{2} + a_{\alpha \beta'} }, \;\;\; \beta = \alpha, \beta' \neq \alpha,\\
&\frac{\varepsilon_{3} + a_{\beta \alpha} }{a_{\beta \alpha} }, \;\;\; \beta \neq \alpha, \beta' = \alpha.
\end{align*}
 where the notation $a_{\alpha \beta} = a_{\alpha} - a_{\beta}$ is used.
If $\beta = \beta'$ then the factor is equal to 
\begin{align*}
\frac{(\varepsilon_{3} + \varepsilon_{2})(\varepsilon_{3} + \varepsilon_{1})}{\varepsilon_{2} \varepsilon_{1}}.
\end{align*}
Thus $Z_{1}^{inst}$ is given by the sum
\begin{align}\label{Z1inst}
Z^{inst}_{1} =  \frac{(\varepsilon_{3} + \varepsilon_{2})(\varepsilon_{3} + \varepsilon_{1})}{\varepsilon_{2} \varepsilon_{1}} \sum\limits_{\alpha = 1}^{N} \prod\limits_{\beta \neq \alpha}\frac{(\varepsilon_{3} + a_{\beta \alpha} )(\varepsilon_{3} + \varepsilon_{1} + \varepsilon_{2} + a_{\alpha \beta} )}{a_{\beta \alpha} (\varepsilon_{1} + \varepsilon_{2} + a_{\alpha \beta} )}.
\end{align}
The second order term has a contributions from three kinds of collections of partitions:
\begin{align*}
&(\lambda^{(\alpha)}) = (\varnothing, ..., \varnothing, \stackrel{\alpha}{\tiny\yng(1,1)}, \varnothing, ..., \varnothing), \\
&(\lambda^{(\alpha)}) = (\varnothing, ..., \varnothing, \stackrel{\alpha}{\tiny\yng(2)}, \varnothing, ..., \varnothing), \\
&(\lambda^{(\alpha)}) = (\varnothing, ..., \varnothing, \stackrel{\alpha_1}{\tiny\yng(1)}, \varnothing, ..., \stackrel{\alpha_2}{\tiny\yng(1)}, \varnothing,...,\varnothing).
\end{align*}
We have
\begin{align*}
Z^{inst}_{\fontsize{2.5}{4}\selectfont\yng(1,1)_{\alpha}} = &\frac{(\varepsilon_3 - \varepsilon_{1} + \varepsilon_{2})(\varepsilon_3 + 2\varepsilon_1 )(\varepsilon_3 + \varepsilon_2 )(\varepsilon_3 + \varepsilon_1 )}{(-\varepsilon_{1} + \varepsilon_{2})  2\varepsilon_1 \varepsilon_2 \varepsilon_1 }  \\
& \prod\limits_{\beta \neq \alpha} \frac{(\varepsilon_3 + 2\varepsilon_1 + \varepsilon_2 + a_{\alpha \beta})(\varepsilon_3 + \varepsilon_1 + \varepsilon_2 + a_{\alpha \beta})(\varepsilon_3 - \varepsilon_1 + a_{\beta \alpha})(\varepsilon_3 + a_{\beta \alpha})}{(2\varepsilon_1 + \varepsilon_2 + a_{\alpha \beta})(\varepsilon_1 + \varepsilon_2 + a_{\alpha \beta})(-\varepsilon_1 + a_{\beta \alpha}) a_{\beta \alpha}},
\end{align*}

\begin{align*}
Z^{inst}_{\fontsize{2.5}{4}\selectfont\yng(2)_{\alpha}} = &\frac{(\varepsilon_3 - \varepsilon_{2} + \varepsilon_{1})(\varepsilon_3 + 2\varepsilon_2 )(\varepsilon_3 + \varepsilon_1 )(\varepsilon_3 + \varepsilon_2 )}{(-\varepsilon_{2} + \varepsilon_{1})  2\varepsilon_2 \varepsilon_1 \varepsilon_2 }   \\
& \prod\limits_{\beta \neq \alpha} \frac{(\varepsilon_3 + 2\varepsilon_2 + \varepsilon_1 + a_{\alpha \beta})(\varepsilon_3 + \varepsilon_2 + \varepsilon_1 + a_{\alpha \beta})(\varepsilon_3 - \varepsilon_2 + a_{\beta \alpha})(\varepsilon_3 + a_{\beta \alpha})}{(2\varepsilon_2 + \varepsilon_1 + a_{\alpha \beta})(\varepsilon_2 + \varepsilon_1 + a_{\alpha \beta})(-\varepsilon_2 + a_{\beta \alpha}) a_{\beta \alpha}},
\end{align*}

\begin{align*}
Z^{inst}_{\fontsize{2.5}{4}\selectfont\yng(1)_{\alpha_1 },  \fontsize{2.5}{4}\selectfont\yng(1)_{\alpha_2 }} =&
\left( \frac{(\varepsilon_3 + \varepsilon_2 )(\varepsilon_3 + \varepsilon_1 )}{\varepsilon_2 \varepsilon_1 }\right)^2 \frac{\left((\varepsilon_3 + \varepsilon_2 )^2 - a_{\alpha_1 \alpha_2}^2\right) \left( (\varepsilon_3 + \varepsilon_1 )^2 - a_{\alpha_1 \alpha_2 }^2\right)}{(\varepsilon_2^2 - a_{\alpha_1 \alpha_2 }^2)(\varepsilon_1^2 - a_{\alpha_1 \alpha_2}^2)} \\
&\prod\limits_{\beta \neq \alpha_1 , \alpha_2} 
\frac{(\varepsilon_3 + \varepsilon_1 + \varepsilon_2 + a_{\alpha_1 \beta})(\varepsilon_3 + \varepsilon_1 + \varepsilon_2 + a_{\alpha_2 \beta})(\varepsilon_3 + a_{\beta \alpha_1})(\varepsilon_3 + a_{\beta \alpha_2 })}{(\varepsilon_1 + \varepsilon_2 + a_{\alpha_1 \beta})(\varepsilon_1 + \varepsilon_2 + a_{\alpha_2 \beta})a_{\beta \alpha_1} a_{\beta \alpha_2 }}.
\end{align*}
The 2-instanton term hence is given by
\begin{align*}
    Z^{inst}_{2} = \sum\limits_{\alpha = 1}^{N} Z^{inst}_{\fontsize{2.5}{4}\selectfont\yng(1,1)_{\alpha}} + Z^{inst}_{\fontsize{2.5}{4}\selectfont\yng(2)_{\alpha}} + \frac{1}{2}\sum\limits_{\alpha_1 \neq \alpha_2} Z^{inst}_{\fontsize{2.5}{4}\selectfont\yng(1)_{\alpha_1 },  \fontsize{2.5}{4}\selectfont\yng(1)_{\alpha_2 }}
\end{align*}

\subsection{Problem \myhref{prob5}{5}}
Expanding
\begin{align*}
    \varepsilon_1 \varepsilon_2 \log \left( 1 + \mathfrak{q} \varepsilon_1 \varepsilon_2 Z^{inst}_{1} + \mathfrak{q}^2 Z^{inst}_{2} + ... \right) =\mathfrak{q} Z^{inst} + \mathfrak{q}^2 \varepsilon_1 \varepsilon_2 (Z^{inst}_2 - \frac{1}{2}Z^{inst 2}_1 ) + ...
\end{align*}
we see that
\begin{align*}
    \mathcal{F}^{inst}_{1} = \varepsilon_1 \varepsilon_2 Z^{inst}_{1} , \;\;\;
    \mathcal{F}^{inst}_{2} = \varepsilon_1 \varepsilon_2 (Z^{inst}_2 - \frac{1}{2}Z^{inst 2}_1 )
\end{align*}
From the formula (\ref{Z1inst}) we easily see
\begin{align*}
    \mathcal{F}_{1}^{inst} = \varepsilon_3^2 \sum\limits_{\alpha} \prod\limits_{\beta \neq \alpha} \frac{\varepsilon_3^2 - a_{\alpha \beta}^2}{a_{\alpha \beta}^2} + O(\varepsilon_1 , \varepsilon_2 )
\end{align*}
To establish the regularity of $\mathcal{F}^{inst}_{2}$ we write it out as
\begin{align*}
    \mathcal{F}^{inst}_{2} = \varepsilon_1 \varepsilon_2 \left(\sum\limits_{\alpha} Z^{inst}_{\fontsize{2.5}{4}\selectfont\yng(1,1)_{\alpha}} + Z^{inst}_{\fontsize{2.5}{4}\selectfont\yng(2)_{\alpha}} -\frac{1}{2} Z^{inst 2}_{\fontsize{2.5}{4}\selectfont\yng(1)_{\alpha}} + \frac{1}{2}\sum\limits_{\alpha_1 \neq \alpha_2} Z^{inst}_{\fontsize{2.5}{4}\selectfont\yng(1)_{\alpha_1 },  \fontsize{2.5}{4}\selectfont\yng(1)_{\alpha_2 }} - Z^{inst}_{\fontsize{2.5}{4}\selectfont\yng(1)_{\alpha_1}} Z^{inst}_{\fontsize{2.5}{4}\selectfont\yng(1)_{\alpha_2}}\right)
\end{align*}
\newpage
The singular part of the first two terms equal to
\begin{align*}
    \frac{\varepsilon_3^4}{2}\left( \frac{1}{\varepsilon_1 (\varepsilon_2 - \varepsilon_1)} + \frac{1}{\varepsilon_2 (\varepsilon_1 - \varepsilon_2 )}\right) \prod\limits_{\beta \neq \alpha} \frac{(\varepsilon_3 + a_{\alpha \beta})^2 (\varepsilon_3 - a_{\alpha \beta})^2}{a_{\alpha \beta}^4} = \frac{\varepsilon_1 \varepsilon_2}{2}\left( \frac{\varepsilon_3^2}{\varepsilon_1 \varepsilon_2} \prod\limits_{\beta \neq \alpha} \frac{\varepsilon_3^2 - a_{\alpha \beta}^2}{a_{\alpha \beta}^2}\right)^2 ,
\end{align*}
which is subtracted with singular term of $\frac{\varepsilon_1 \varepsilon_2 }{2} Z^{inst 2}_{\fontsize{2.5}{4}\selectfont\yng(1)_{\alpha}}$. Also the singular part of $\varepsilon_1 \varepsilon_2 Z^{inst}_{\fontsize{2.5}{4}\selectfont\yng(1)_{\alpha_1 },  \fontsize{2.5}{4}\selectfont\yng(1)_{\alpha_2 }}$
\begin{align*}
    \varepsilon_1 \varepsilon_2 \left( \frac{\varepsilon_3^2 (\varepsilon_3^2 - a_{\alpha_1 \alpha_2}^2)^2}{\varepsilon_1 \varepsilon_2 a_{\alpha_1 \alpha_2}^2}\right)^2 \prod\limits_{\beta \neq \alpha_1 , \alpha_2} \frac{(\varepsilon_3^2 - a_{\alpha_1 \beta}^2)(\varepsilon_3^2 - a_{\alpha_2 \beta}^2)}{a_{\alpha_1 \beta}^2 \alpha_{\alpha_2 \beta}^2}
\end{align*}
coincides with the one of $\varepsilon_1 \varepsilon_2 Z^{inst}_{\fontsize{2.5}{4}\selectfont\yng(1)_{\alpha_1}} Z^{inst}_{\fontsize{2.5}{4}\selectfont\yng(1)_{\alpha_2}}$, so all the singular terms vanish.
\subsection{Problem \myhref{prob6}{6}}
Let us formally compute the expectation value of the product of some local operator $\CalO(\varphi)$ with non-regularized operator $e^{2\pi \ii n {\varphi}(x)}$ in case of $M^{3} =  \mathbb{R}^{3}$ with the standard Euclidean metric:
\begin{align}\label{expvalver}
    \langle \CalO(\varphi) e^{2\pi \ii n {\varphi(x)}} \rangle = \frac{1}{Z}\int [D{\varphi}] \, \CalO(\varphi) e^{2\pi \ii n {\varphi}(x) -g^2 \int_{M^3} d{\varphi} \wedge \star d{\varphi}},
\end{align}
where $Z$ is the partition function
\begin{align*}
    Z = \int [D{\varphi}] \, e^{ - g^2 \int_{M^3} d{\varphi} \wedge \star d{\varphi}}.
\end{align*}
Consider Green's function $K(x, x') = \frac{1}{|x - x'|}$ for the Laplace operator:
\begin{align*}
     \Delta_{x'}K(x, x') = 4 \pi \delta^{(3)}_{x}(x').
\end{align*}
The exponent in (\ref{expvalver}) can be represented in the following way
\begin{align*}
    &2\pi \ii n {\varphi}(x) -g^2 \int_{M^3} d{\varphi} \wedge \star d{\varphi}=\\
    &=\frac{\ii n}{2} \int_{\mathbb{R}^3} {\varphi}(x') \Delta_{x'} K(x, x') d^{3} x' -g^2 \int_{\mathbb{R}^3} {\varphi}(x') \Delta {\varphi}(x') d x' = \\
    &=-g^2 \int_{\mathbb{R}^3} \left({\varphi}(x') - \frac{\ii n}{4g^{2}} K(x, x')\right) \Delta \left({\varphi}(x') - \frac{\ii n}{4g^{2}} K(x, x')\right) - \frac{n^2}{16g^2} K(x, x)
\end{align*}
Shifting variable in path integral
\begin{align*}
   {\varphi}(x') \rightarrow {\varphi}(x') - \frac{in}{4g^{2}} K(x, x')
\end{align*}
we obtain
\begin{align*}
     \langle \CalO(\varphi) e^{2\pi \ii n {\varphi(x)}} \rangle = \frac{1}{Z} \int [D{\varphi}] \, \CalO(\varphi') e^{ -g^2 \int_{M^3} d{\varphi} \wedge \star d{\varphi}} e^{-\frac{n^{2}}{16g^2}K(x, x)} \sim e^{-\frac{n^{2}}{16g^2}K(x, x)},
\end{align*}
where $\varphi'$ is shifted $\varphi$.
Of course, $K(x, x)$ is not defined, but after regularization we get
\begin{align*}
     \langle \CalO(\varphi) e^{2\pi \ii n \frac{1}{A(S^{2}_{\epsilon}(x))}\int_{S^{2}_{\epsilon}(x)}{\varphi(x')}dx'} \rangle \sim e^{-\frac{n^{2}}{16g^2}\frac{1}{A(S^2_{\epsilon}(x))}\int_{S^{2}_{\epsilon}(x)}K(x, x')dx'} = e^{-\frac{1}{16}\frac{n^2}{g^2\epsilon}}.
\end{align*}
Thus, the regularized operator should be defined as
\begin{align*}
    {\CalO}_{n}(x) = {\rm Lim}_{{\ep} \downarrow 0} \ e^{2\pi \ii n \int_{S^{2}_{\ep}(x)} {\varphi}(x') d^2x'}\ e^{\frac{1}{16} \frac{n^2}{g^2 \ep}}.
\end{align*}
\subsection{Problem \myhref{prob7}{7}}
To investigate $U(1)$-valued sigma model counterpart of the Wilson loop in abelian $U(1)$ Maxwell theory it is convenient to introduce path integral for a coupled system of gauge $U(1)$ field $A$ and unconstrained $1$-form $\alpha \in \Omega^{1}(M^{3})$
\begin{align}\label{corralph}
    \int[DA][D\alpha] \, e^{-g^{2}\int_{M^{3}}\alpha \wedge \star \alpha + \ii \int_{M^{3}}F \wedge \alpha}.
\end{align}
Note that this is a different system compared to the considered (\ref{eq:max3d2}).

Integrating out $\alpha$, we get the effective Lagrangian
\begin{align*}
    \int[DA] \, e^{-\frac{1}{4g^{2}}\int_{M^{3}}F \wedge \star F }.
\end{align*}
On the other hand, integrating out $A$, summing over possible values of the first Chern class of $U(1)$-bundle $[\frac{F}{2\pi}] \in H^{2}(M^{3}, \mathbb{Z})$, we get the constraints: $d\alpha = 0$, and the cohomology class of $\alpha$ is integral:
\begin{align*}
    [\alpha] \in H^{1}(M, \mathbb{Z}).
\end{align*}
Thus, $\alpha = d {\varphi}$ for some $U(1)$-valued scalar field ${\varphi}$, so we get the effective Lagrangian of $U(1)$-valued sigma model
\begin{align*}
     \int[D{\varphi}] \, e^{-g^{2}\int_{M^{3}}d{\varphi} \wedge \star d{\varphi} }.
\end{align*}
Now consider (\ref{corralph}) with (\ref{abwil}) inserted:
\begin{align*}
     &\int[DA][D\alpha] \, e^{-g^{2}\int_{M^{3}}\alpha \wedge \star \alpha + \ii \int_{M^{3}}F \wedge \alpha} P e^{\ii n \oint_{C}A} 
     =\int[DA][D\alpha] \, e^{-g^{2}\int_{M^{3}}\alpha \wedge \star \alpha - \ii \int_{M^{3}}A \wedge d\alpha + \ii n \int_{M^{3}}A \wedge \delta_{C}}=\\
     &=\int[DA][D\alpha] \, e^{-g^{2}\int_{M^{3}}\alpha \wedge \star \alpha - \ii \int_{M^{3}}A \wedge (d\alpha -n \delta_{C}) },
\end{align*}
where $\delta^{(2)}_{C}$ is a two-form with delta-function support along the loop $C$. Integrating over $A$ we get, instead of the constraint $d\alpha = 0$, the constraint
\begin{align*}
    d\alpha = n \delta^{(2)}_{C}.
\end{align*}
In other words, the $1$-form $\alpha$ has a nontrivial circulation $n$ around arbitrary small circle bridged with the loop $C$. This means that the $U(1)$-valued field $\varphi$ is well-defined on $M^{3} \setminus C$, it has winding number $n$ along a small circle bridged with $C$.  
\subsection{Problem \myhref{prob8}{8}}
Let $A$ be a diagonal matrix $A = \mathrm{diag}(a_{0}, ..., a_{N-1})$.  The matrix $1 - z^{-1} A$ has the form
\begin{align}\label{1-za}
    1 - C_{z^{-1}} A =  \begin{pmatrix}
        1 & -a_{1}\\
          & 1 & -a_{2}\\
          &   & \cdot & \cdot\\
          &   &  & \cdot & \cdot\\
          &   &  &  & 1 & -a_{N-1}\\
         -z^{-1}a_{0} &  &  &  &      & 1  
    \end{pmatrix}.
\end{align}
Expanding the determinant by the first column we get
\begin{align*}
    &\det (1 - C_{z^{-1}} A) = 1 + (-1)^{N-1}(-z^{-1}a_{0})\det
    \begin{pmatrix}
        -a_{1} & \\
         1 & -a_{2} & \\
          &  1 & \cdot & \cdot\\
          &   &  & \cdot & \cdot\\
          &   &  &  1 & -a_{N-2} & \\
          &  &  &  &   1   & -a_{N-1}  
    \end{pmatrix}=\\
    &=1 - z^{-1}a_{0}\cdot ... \cdot a_{N_1} = 1 - z^{-1}\det A.
\end{align*}
\subsection{Problem \myhref{prob9}{9}}
Consider the operator $ \hat{D}_{0} = \prod\limits_{i = 0}^{\overset{\longrightarrow}{r}}\big( 1 - \hat{C}_{z} \hat{Z}_{i}\big)$. It degenerates when $\det (1 - z^{-1}\hat{C}_{z} \hat{Z}_{i})=0$ for some $i$. By the previous problem it happens when $z = \det \hat{Z}_{i} = z_{i}$. Since $z_{i} \neq z_{j}$ for different $i$ and $j$ the operators $1 - \hat{C}_{z_i}\hat{Z}_{j}$ are invertible and thus the dimension of $\Ker \hat{D}_{0}(z_{0})$ is equal to the dimension of $\Ker (1 - \hat{C}_{z_i} \hat{Z}_{i})$. From the explicit form (\ref{1-za}) we see that the first $N-1$ rows of  $1 - \hat{C}_{z} \hat{Z}_{i}$ are always independent so $\mathrm{rk}\, (1 - \hat{C}_{z_0} \hat{Z}_{i}) = N-1$ and hence $\dim\Ker(1 - \hat{C}_{z} \hat{Z}_{i}) = 1$. Now consider the product
\begin{align*}
    \hat{D}_{0}(z)\hat{L}(z) = -\hat{D}_{1}(z) \underset{z \rightarrow z_{i}}{=} O(1).
\end{align*}
Expanding the lhs we get
\begin{align*}
    \left(\hat{D}_{0}(z_0 ) + O(z-z_{0})\right)\left(\frac{\mathrm{Res}_{z = z_0} \hat{L}(z)}{z - z_{0}} + O(1)\right) = \frac{\hat{D}_{0}(z_0 ) \mathrm{Res}_{z = z_0}\hat{L}(z)}{z - z_{0}} + O(1).
\end{align*}
Since the rhs is regular we conclude $\hat{D}_{0}(z_0 ) \, \mathrm{Res}_{z = z_0}\hat{L}(z) = 0$. In other words 
\begin{align*}
    \mathrm{im}\, \mathrm{Res}_{z = z_0}\, \hat{L}(z) \subset \Ker \hat{D}_{0}(z_0 ).
\end{align*}
It follows that $\dim\, \mathrm{im}\, \mathrm{Res}_{z = z_0}\hat{L}(z)$ is no greater than $1$, and since $\hat{L}(z)$ is singular $\mathrm{Res}_{z = z_0}\hat{L}(z) \neq 0$, so $\mathrm{rk}\,  \mathrm{Res}_{z = z_0}\hat{L}(z) = 1$.

\newpage
\newcommand{\etalchar}[1]{$^{#1}$}

\end{document}